\documentclass[twocolumn,aps,prl,superscriptaddress,longbibliography]{revtex4-1}
\usepackage[colorlinks,bookmarks=false,citecolor=blue,linkcolor=red,urlcolor=blue]{hyperref}
\usepackage{verbatim}
\usepackage{color}
\usepackage{amsmath,amssymb,bm,braket,float,mathtools,subfigure}
\usepackage{graphicx,xfrac,appendix}
\usepackage[english]{babel}

\makeatletter

\usepackage{epstopdf,float}

\makeatother
\begin{document}

\title{Fate and origin of the quantum Otto heat  engine based on the dissipative Dicke-Hubbard model}

\author{He-Guang Xu}
\affiliation{School of Optoelectronic Engineering, Guangdong Polytechnic Normal University, Guangzhou 510665, China}

\author{Shujie Cheng}
\email{chengsj@zjnu.edu.cn}
\affiliation{Xingzhi College, Zhejiang Normal University, Lanxi 321100, China}
\affiliation{Department of Physics, Zhejiang Normal University, Jinhua 321004, China}

\date{\today}

\begin{abstract}
The Dicke-Hubbard model, describing an ensemble of interacting atoms in a cavity, provides a rich platform for exploring 
collective quantum phenomena. However, its potential for quantum thermodynamic applications remains largely uncharted. Here, 
we study a quantum Otto heat engine whose working substance is a system governed by the Dicke-Hubbard 
Hamiltonian. Through the research on steady-state superradiance phase transitions, it is demonstrated that the 
steady-state synergistic mechanism under high and low temperature environments is the reason for the emergence of 
high-performance heat engines. By analyzing the influences of atom-light coupling strength, inter-cavity hopping strength 
and atom number on the working modes of quantum Otto cycle,  it is clarified that the effective working regions of each 
working mode. This work has established a close connection between superradiance phase transition and the 
quantum thermodynamic applications. It not only deepens our understanding of the energy conversion mechanism in 
non-equilibrium quantum thermodynamics but also lays a theoretical foundation for the future experimental design of 
high-performance quantum Otto heat engines.
\end{abstract}

%\pacs{}
%keywords

\maketitle

\section{Introduction}

The burgeoning field of quantum thermodynamics seeks to extend the principles of classical thermodynamics to the quantum realm, revealing novel mechanisms for work extraction, refrigeration, and energy transduction~\cite{Vinjanampathy2016,Millen2016}. Among various paradigms, the quantum Otto cycle stands as a fundamental and experimentally feasible model for a quantum heat engine~\cite{Kosloff2013,Quan2007}. By alternately coupling a quantum working substance to a hot and a cold bath during an isochoric process and modulating its Hamiltonian adiabatically, the cycle converts heat into useful work. The performance of such an engine is intrinsically tied to the quantum properties of the working substance—such as coherence, entanglement, and quantum criticality~\cite{Uzdin2015,Campisi2016}. 

Early theoretical and experimental studies have explored various systems as working substances.  In the aspect of theoretical research, 
it involves single spins~\cite{Zhang2014,feldmann2000performance,geva1992quantum}, bosonic substances \cite{deAssis_2021}, harmonic oscillators ~\cite{Rossnagel2014,abah2012single,myers2020bosons}, and few-level atoms ~\cite{Klaers2017,2012Quantum}, coupled spin systems~\cite{2004Characteristics,huang2013special,huang2014specialf}, coupled spin-$3/2$~\cite{ ivanchenko2015quantum}, relativistic oscillators~\cite{Myers_2021}, Bose-Einstein condensates~\cite{wang2009performancef,myers2022boostingf}, and light-matter systems described by the Jaynes-Cumming ~\cite{altintas2015rabi,song2016one,barrios2017role,mojaveri2021quantumf}, quantum Rabi~\cite{jaynes1963comparison,braak2011integrability,chen2012exact,braak2016semi,rabi1936process,rabi1937space} models, anisotropic quantum Rabi-Stark model~\cite{xu2024exploring} and  Dicke model~\cite{xu2024universal}.  In the aspect of experimental research, the quantum heat engine are realized in the platforms including 
superconducting circuits~\cite{quan2006maxwell,niskanen2007information,pekola2010decoherence,koski2015chip,pekola2016maxwell}, trapped ion systems~\cite{abah2012single,rossnagel2014nanoscale,rossnagel2016single,maslennikov2019quantumf}, optomechanics~\cite{zhang2014quantum,2014Theory}, ultracold atoms~\cite{2012Isolated,brantut2013thermoelectric}, nuclear magnetic resonance ~\cite{batalhao2014experimental,micadei2019reversing, de2019efficiency}. 

While earlier findings provide foundational insights, the enhanced power output or unique quantum advantages 
caused by many-body physics are not fully demonstrated. Recently, there has been a growing interest in employing many-body 
systems, such as ultracold atoms in optical lattices~\cite{Bouton2021} and interacting spin chains~\cite{Alsulami2024tow}, where 
phenomena like superradiance and the Bose-Hubbard phase transition have been shown to significantly boost engine 
performance or induce new operational modes~\cite{hardal2015superradiant,Jaramillo2016}. Furthermore, the Dicke model incorporates 
both many-body and superradiance physics as well, which describes an ensemble of two-level atoms collectively coupled to a single-mode cavity field~\cite{Dicke1954}. This model exhibits a well-known superradiant quantum phase transition which can serve as a potent resource for quantum thermodynamics~\cite{Bhaseen2012}. On the other hand, the Bose-Hubbard model, central to describing strongly correlated bosons in lattices, features a Mott insulator-to-superfluid transition ~\cite{Greiner2002}. The synthesis of these two models—the Dicke-Hubbard (DH) model~\cite{PhysRevLett115180404,Nataf2010}—creates a rich landscape where atomic hopping, on-site interactions, and collective light-matter coupling compete and intertwine. This hybrid system offers a unique testbed for investigating how the interplay between optical coherence and matter-wave correlations influences energy conversion processes.

Although extensive research has been conducted on the equilibrium and non-equilibrium properties of the DH model, 
its potential as the working substance of quantum heat machine has not been largely explored. Crucially, the roles of the 
superradiance phase transition in thermodynamic cycles such as the Otto cycle remain unclear. 
The key questions remain: How does collective photon-atom interaction improve work efficiency and effectiveness?  
What is the origin of the quantum heat engine?

This paper is organized as follows. In Sec. \ref{S2} we introduce the DH model, the extended bosonic coherent state approach, 
and the quantum dressed master equation. 
 In Sec. \ref{S3} we introduce the quantum Otto cycle. In Sec. \ref{S4}, we present the results and make discussions. A conclusion is made in 
 Sec. \ref{S5}.

\section{Model and Method}\label{S2}
\subsection{The Dicke-Hubbard model}

The Hamiltonian describing the DH model  consisting of a single-mode light field interacting with $N$ identical two-level atoms and photon hopping between the nearest-neighboring cavity with the strength $J$, is expressed as~\cite{greentree2006quantum,hartmann2006strongly,hartmann2008quantum,lei2008quantum} ($\hbar=1$) 

\begin{equation}~\label{eq1}
\hat{H}=\sum_{i}\hat{H_{i}}^{Dicke}-J\sum_{\langle i,j\rangle}\hat{a}_{i}^{\dagger}\hat{a}_{j}
\end{equation}
where $\hat{H_{i}}^{Dicke}$ denotes the Hamiltonian of the Dcike model at $i$th site, and is given by
($\hbar=1$)~\cite{lu2016influencef,kirton2019introduction}
\begin{equation}~\label{h0}
\hat{H_{i}}^{Dicke}={\omega_{0}}\hat{a}^{\dagger}_{i}\hat{a}_{i}+\Delta \hat{J}_{z}^{i}+\frac{2\lambda}{\sqrt{N}} (\hat{a}^{\dagger}_{i}+\hat{a}_{i})\hat{J}_{x}^{i},
\end{equation}
where $\omega_{0}$ and $\Delta$ are the frequencies of the single-mode and atoms, respectively, $\lambda$ is the atom-light coupling strength, 
$\hat{a}^{\dagger}_{i}$($\hat{a}_{i}$) denotes the creation (annihilation) operator of the bosonic field at $i$th site, $\hat{J}_{x}^{i}=\frac{1}{2}(\hat{J}_{+}^{i}+\hat{J}_{-}^{i})$ and $\hat{J}_{z}^{i}$ are the pseudospin operators at $i$th site given by $\hat{J}_{{\pm}}^{i}=\text{\ensuremath{\sum}}_{k}^{N}\hat{\sigma}_{\pm}^{i,k},\hat{J}_{z}^{i}=\sum_{k}^{N}\hat{\sigma}_{z}^{i,k}$, with $\hat{\sigma}_\alpha~(\alpha=x,y,z)$ being the Pauli operators.
The pseudospin operators satisfy the commutation relation
$[\hat{J}_{+}^{i},\hat{J}_{-}^{i}]=2\hat{J}_{z}^{i}$ , $[\hat{J}_{z}^{i},\hat{J}_{\pm}^{i}]=\pm \hat{J}_{\pm}^{i}$.,$J$ describes the interaction strength 
of the inter-cavity hopping. The Dicke model has been extensively studied due to its ability to undergo superradiance quantum phase 
transition at critical point $\lambda_{c}=\sqrt{\omega\Delta}/2$. The model preserves the parity symmetry with 
parity operator $\hat{P}=\Pi_{i}\hat{P}_{i}$, where $\hat{P}_{i}=\exp\{i\pi\hat{\Lambda}_{i}\}$ is the parity operator of the $i$th site, and $\hat{\Lambda}_{i}=\hat{J}_{z}^{i}+\hat{a}^{\dagger}_{i}\hat{a}_{i}$ is the total excitation number.

\subsection{Extended bosonic coherent state approach}
We can transform the Dicke-Hubbard model to an effective single-site Dicke model by using the mean-field approximation. 
Specifically, the inter-site photon hopping term in  Eq.~(\ref{eq1}) is decoupled as $\hat{a}_{i}^{\dagger}\hat{a}_{j}=\psi^{*}\hat{a}_{i}+\psi\hat{a}^{\dagger}_{j}-|\psi|^{2}$, with $\psi=\langle\hat{a}\rangle$ the superfluid order parameter. Therefore, the $i$th site Hamiltonian is given by
\begin{equation}
 \hat{H}_{i}=\hat{H_{i}}^{Dicke}-zJ(\psi\hat{a}_{i}^{\dagger}+\psi^{*}\hat{a}_{i})+zJ|\psi|^{2}
\end{equation}
where $z$ is the number of nearest-neighbor sites, we set $z=3$ for two-dimensional cavity lattice\cite{hartmann2006strongly,schiro2013quantum,lu2016influence} in this paper.  Since real order parameter is considered, i.e., $\psi=\psi^{*}$, 
the reduced model becomes site independent, resulting in the effective mean-field Hamiltonian
\begin{equation}~\label{mf}
\hat{H}_{MF}={\omega_{0}}\hat{a}^{\dagger}\hat{a}+\Delta \hat{J}_{z}+\frac{2\lambda}{\sqrt{N}} (\hat{a}^{\dagger}+\hat{a})\hat{J}_{x}-zJ\psi(\hat{a}^{\dagger}+\hat{a})+zJ|\psi|^{2}
\end{equation}

Dicke model has numerically exact solution by using extended bosonic coherent state approach ~\cite{chen2008numerically,chen2010quantum}, 
which is used to obtain the numerical solution of Eq.~(\ref{mf}) in a self-consistent way as well. 
At first, we introduce the angular momentum operator $\hat{J}_y$ to perform a rotation transformation on $\hat{H}_{MF}$, 
i.e., $\hat{H}^{s}_{MF}=\exp(i{\pi}\hat{J}_y/2)\hat{H}_{MF}\exp(-i{\pi}\hat{J}_y/2)$. Thus, we have 
\begin{eqnarray}~\label{mff}
\hat{H}^{s}_{MF}&=&\omega_{0} \hat{a}^{\dagger}\hat{a}-\frac{\Delta}{2}(\hat{J}_{+}+\hat{J}_{-})\nonumber\\
&+&(\hat{a}^{\dagger}+\hat{a})(\frac{2\lambda}{\sqrt{N}}\hat{J}_{z}-zJ\psi)+zJ\psi^{2}.
\end{eqnarray}
The corresponding wavefunction of $\hat{H}^{s}_{MF}$ can be expressed in terms of
the basis $\{|\varphi_{m}{\rangle}_{b}\otimes|j,m{\rangle}\}$, $\{|j,m{\rangle},m=-j,-j+1,...,j-1,j$\} ($j=N/2$) is the 
Dicke state for the two-level atoms. By considering the displacement transformation $\hat{A}_{m}=\hat{a}+g_{m}$ with $g_{m}=\frac{2\lambda m}{\omega_{0}\sqrt{N}}-\frac{zJ\psi}{\omega_{0}}$, 
the Schr\"{o}dinger equation of $\hat{H}^{s}_{MF}$ on the basis is given by
\begin{eqnarray}\label{Schr}
&&-\Delta j_{m}^{+}|\varphi_{m}{\rangle}_{b}|j,m+1{\rangle}-\Delta j_{m}^{-}|\varphi_{m}{\rangle}_{b}|j,m-1{\rangle}\nonumber\\
&&+\omega_{0}(\hat{A}_{m}^{\dagger}\hat{A}_{m}-g_{m}^{2}+zJ\psi^{2})|\varphi_{m}{\rangle}_{b}|j,m{\rangle}\nonumber\\
&&=E|\varphi_{m}{\rangle}_{b}|j,m{\rangle},
\end{eqnarray}
where $\hat{J}_{\pm}|j,m{\rangle}=j_{m}^{\pm}|j,m\pm 1{\rangle}$, with $j_{m}^{\pm}=\frac{1}{2}\sqrt{j(j+1)-m(m\pm1)}$.
Next, we multiply Eq.~(\ref{Schr}) on the left by $\{{\langle}n,j|\}$, which results in 
\begin{eqnarray}
&&-\Delta j_{n}^{+}|\varphi_{n+1}{\rangle}_{b}-\Delta j_{n}^{-}|\varphi_{n-1}{\rangle}_{b}\nonumber\\
&&+\omega_{0}(\hat{A}_{n}^{\dagger}\hat{A}_{n}-g_{n}^{2}+zJ\psi^{2})|\varphi_{n}{\rangle}_{b}=E|\varphi_{n}{\rangle}_{b},
\end{eqnarray}
where $n=-j,-j+1,...,j$.
Furthermore, the bosonic state can be expanded as
\begin{eqnarray}
|\varphi_{m}{\rangle}_{b}
&=&\sum_{k=0}^{\textrm{N}_\textrm{tr}}\frac{1}{\sqrt{k!}}
c_{m,k}(\hat{A}_{m}^{\dagger})^{k}|0{\rangle}_{A_{m}}\nonumber\\
&=&\sum_{k=0}^{\textrm{N}_\textrm{tr}}\frac{1}{\sqrt{k!}}
c_{m,k}(\hat{a}^{\dagger}+g_{m})^{k}e^{-g_{m}
\hat{a}^{\dagger}-g_{m}^{2}/2}|0{\rangle}_{a},
\end{eqnarray}
where $\textrm{N}_\textrm{tr}$ is the truncation number of photonic excitations.
Finally, we obtain the eigenvalue equation
\begin{eqnarray}
&&\omega_{0}(l-g_{n}^{2}+zJ\psi^{2})c_{n,l}-\Delta j_{n}^{+}\sum_{k=0}^{\textrm{N}_\textrm{tr}}
c_{n+1,kA_{n}}{\langle}l|k{\rangle}_{A_{n+1}}\nonumber\\
&&-\Delta j_{n}^{-}\sum_{k=0}^{\textrm{N}_\textrm{tr}}
c_{n-1,kA_{n}}{\langle}l|k{\rangle}_{A_{n-1}}=Ec_{n,l}.~
\end{eqnarray}
where the coefficients are $_{A_{n}}{\langle}l|k{\rangle}_{A_{n-1}}=(-1)^{l}D_{l,k}$ and
$_{A_{n}}{\langle}l|k{\rangle}_{A_{n+1}}=(-1)^{k}D_{l,k}$,
with
\begin{eqnarray}
D_{l,k}=e^{-G^{2}/2}\sum_{r=0}^{\min[l,k]}
\frac{(-1)^{-r}\sqrt{l!k!}G^{l+k-2r}}{(l-r)!(k-r)!r!},
\end{eqnarray}
where $G=\frac{2\lambda}{\omega_{0}\sqrt{N}}$. 
In the following calculations, we select the truncation number $\textrm{N}_\textrm{tr}=50$, 
which is sufficient to give the convergent excited state energies with relative error less than
$10^{-5}$.  The behavior of the ground-state quantum phase transition of  DH model was studied in Ref.~\cite{lu2016influence} by using an extended coherent state approach within mean-field theory.  Ref.~\cite{ye2021quantum} investigated the quantum phase transition of light in the dissipative Rabi-Hubbard lattice under the framework of the mean-field theory.

\subsection{Quantum dressed master equation}
We study the dissipative DH model where the mean-field Dicke 
model for the $i$th cavity is coupled to two individual bosonic thermal baths with local dissipations. Consequently, 
the total Hamiltonian $\hat{H}_{tot}$ under the mean-field approximation can be expressed as
\begin{equation} 
\hat{H}_{tot}=\hat{H}_{MF}^{s}+\hat{H}_{B}+\hat{V}. 
\end{equation}

Here, the Hamiltonian of the thermal baths $\hat{H}_B$ is expressed as,
\begin{equation}
\hat{H}_B=\sum_{u=q,c}\sum_{k}\omega_k\hat{b}^\dag_{u,k}\hat{b}_{u,k},
\end{equation}
where $\hat{b}^\dag_{u,k}~(\hat{b}_{u,k})$ creates (annihilates) one phonon in the $u$th bath with the frequency $\omega_k$.
The interactions between the Dicke system with thermal baths $\hat{V}$ is specified as
\begin{equation}
\hat{V}=\hat{V}_c+\hat{V}_q,
\end{equation}
with
\begin{eqnarray}
\hat{V}_q&=&\sum_{k}(\lambda_{q,k}\hat{b}^\dag_{q,k}+\lambda^{*}_{q,k}\hat{b}_{q,k}){(\hat{J}_++\hat{J}_-)}/{\sqrt{N}},\\
\hat{V}_c&=&\sum_{k}(\lambda_{c,k}\hat{b}^\dag_{c,k}+\lambda^{*}_{c,k}\hat{b}_{c,k})(\hat{a}^\dag+\hat{a}),
\end{eqnarray}
where $\lambda_{q,k}~(\lambda_{c,k})$ the coupling strength between the atoms  (photons) and the corresponding bath. {The $u$th thermal bath is characterized by the spectral function
$f_{q(c)}(\omega)=2\pi\sum_k|\lambda_{q(c),k}|^2\delta(\omega-\omega_k)$.
In this paper, we specify $f_{q(c)}(\omega)$ the Ohmic case  to quantify the thermal bath, i.e., 
 $f_{q}(\omega)=\gamma_q\omega/\Delta\exp(-\omega/\omega_{co})$, $f_{c}(\omega)=\gamma_c\omega/\omega_0\exp(-\omega/\omega_{co})$,
where $\gamma_{q(c)}$ is the dissipation strength 
 and $\omega_{co}$ is the cutoff frequency of thermal baths.} $\omega_{co}$ is considered to be large enough, so the spectral functions are simplified as  $f_{q}(\omega)=\gamma_q\omega/\Delta$, $f_{c}(\omega)=\gamma_c\omega/\omega_0$.

We consider the case where the interaction between the Dicke system and thermal baths is weak. 
Under the Born-Markov approximation,
the quantum dressed master equation used for investigating the dissipative dynamics of the density matrix  
is given by~\cite{le2016fate,settineri2018dissipation,beaudoin2011dissipation,ye2021quantum,gardiner1985input}
\begin{eqnarray}~\label{eq16}
\frac{d}{dt}\hat{\rho}_s&=&-i[\hat{H}_{MF}^{s},\hat{\rho}_s]+\sum_{u;m<n}
\{\Gamma^{nm}_un_u(\Delta_{nm})\mathcal{D}[|\phi_n{\rangle}{\langle}\phi_m|,\hat{\rho}_s]\nonumber\\
&&+\Gamma^{nm}_u[1+n_u(\Delta_{nm})]\mathcal{D}[|\phi_m{\rangle}{\langle}\phi_n|,\hat{\rho}_s]\}
\end{eqnarray}
where $|\phi_m{\rangle}$ is the eigenfunction of the Dicke-Hubbard model under mean-field theory$\hat{H}_{MF}^{s}$ as $\hat{H}_{MF}^{s}|\phi_m{\rangle}=E_m|\phi_m{\rangle}$, $\Delta_{nm}=E_n-E_m$ is the transition
frequency of two energy levels, 
the dissipator is
$\mathcal{{D}}[\hat{O},\hat{\rho}_s]=\frac{1}{2}[2\hat{O}\hat{\rho}_s\hat{O}^{\dag}-\hat{\rho}_s\hat{O}^{\dag}\hat{O}-\hat{O}^{\dag}\hat{O}\hat{\rho}_s]$,
the  dissipative rates is
$\Gamma^{jk}_u=f_u(\Delta_{nm})|S^{nm}_u|^2=\frac{\gamma_u\Delta_{nm}}{\omega_u}|S^{nm}_u|^2$,
with $\omega_c=\omega_0$, $\omega_q=\Delta$, ${S}^{nm}_q=\frac{1}{\sqrt{N}}{\langle}\phi_n|(\hat{J}_++\hat{J}_-)|\phi_m{\rangle}$
and ${S}^{nm}_c={\langle}\phi_n|(\hat{a}^{\dag}+\hat{a})|\phi_m{\rangle}$.

In the eigenbasis, the dynamics of the population $P_{k}=\langle \phi_{k}|\hat{\rho}_{s}|\phi_{k}\rangle$  is given by
\begin{eqnarray}
\frac{d}{dt}P_{k}
&=&\sum_{u,k{\neq}m}\Gamma^{km}_un_u(\Delta_{km})P_{m}\nonumber\\
&&-\sum_{u,k{\neq}m}\Gamma^{km}_u[1+n_u(\Delta_{km})]P_{k},
\end{eqnarray}
where $\Gamma^{km}_u=-\Gamma^{mk}_u$.

The steady state of the DH model can be self-consistently solving the quantum dressed master equation. The specific process is to 
give an arbitrary reasonable initial value to the order parameter and find a temporary steady state $\rho_{ss}$. And the order 
parameter is $\psi={\rm Tr}\{\rho_{ss}a\}$, for the next-step iteration. This procedure can be repeated until the converged steady state 
and order parameter are achieved. All physical quantities can be calculated by the steady state.

\section{QUANTUM OTTO CYCLE}\label{S3}

%==========================================
\begin{figure}[tbp]
\begin{center}
%\vspace{-2.2cm}
\includegraphics[scale=0.15]{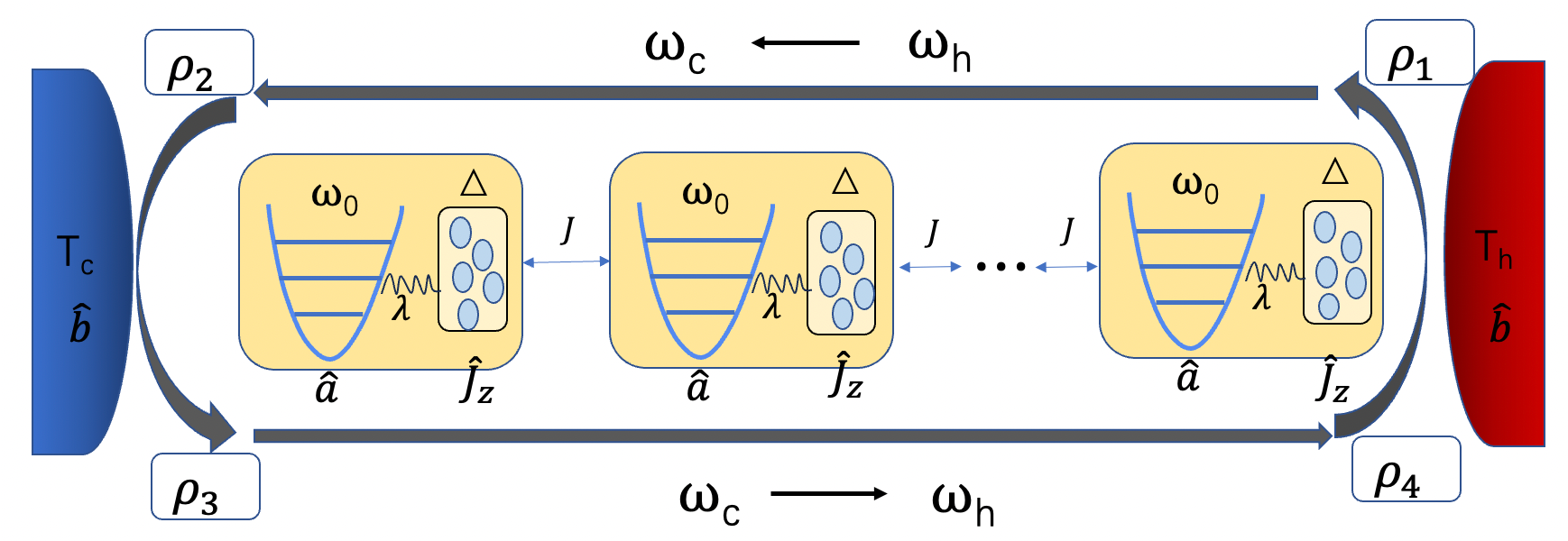}
%\vspace{-1.0cm}
\end{center}
\caption{Schematic representation of the four strokes of an Otto cycle for the realization of a  heat machine based on the open DH model, as detailed in Section III. During the isochoric stroke the frequency of the working substance, as modelled by the DH Hamiltonian, is held fixed while  interacting with  a hot (cold) reservoir at temperature $T_{h}$  ($T_{c}$). Only heat is exchanged during this stroke. In the two quantum adiabatic strokes the working substance is isolated from the reservoir and has its frequency shifted, thus producing work. No heat is exchanged during this stroke. By controlling the parameters $\omega_{0}$, $\Delta$, $J$ and $\lambda$ of the model the machine can work as an engine, refrigerator, heater, or accelerator.
}~\label{fig001}
\end{figure}
%==========================================

We employ the DH model as the working substance to engineer the four-stroke quantum Otto cycle, 
which is composed of two adiabatic and two isochoric processes~\cite{quan2007quantum,quan2009quantum}. 
During the isochoric process, the substance interacts with a hot (cold) reservoir at temperature $T_{h}$ ($T_{c}$ ), and 
the four-stroke quantum Otto cycle has been shown in Fig.~\ref{fig001}). The four strokes are respectively: 

1. Quantum isochoric process. The working substance as modelled by the mean-field Dicke Hamiltonian
$H_{MF,h}^{s}$ with frequency $\omega=\omega_{h}$ ($\omega=\omega_0=\Delta$) 
is brought into contact with a hot reservoir at temperature $T_{h}$. In this process,
the system undergoes a Markovian evolution, which is described by Eq.~(\ref{eq16}).

After a long enough evolution, the system will reach the only steady
state $\rho_{1}=\text{\ensuremath{\rho_{ss}}}(T_{h})=\sum_{n}P_{n}^{ss}(T_{h})|E_{n}^{h}\rangle\langle E_{n}^{h}|$ with
$\frac{d\rho}{dt}=0$, where $P_{n}^{ss}(T_{h})$ is the corresponding population. The eigenstates $|\phi_{k}^{h}\rangle$ and eigenvalues
$E_{k}^{h}$ of $H_{MF,h}^{s}$ are obtained by using the extended bosonic
coherent state approach method \cite{chen2008numerically}. During this process, heat $Q_{h}$ is absorbed from the hot
reservoir, without any work being done.

2. Quantum adiabatic expansion process. The system is isolated from
the hot reservoir and the energy levels is changed from $E_{n}^{h}$
to $E_{n}^{c}$ by varying the frequency from $\omega_{h}$ to $\omega_{c}$
(with $\omega_{h}>\omega_{c}$). This process must be done slow enough
to ensure that the populations $P_{n}^{ss}$ ($T_{h}$) remain unchanged
according to the quantum adiabatic theorem. At the end of this adiabatic expansion, 
the state becomes $\rho_{2}=\sum_{n}P_{n}^{ss}(T_{h})|E_{n}^{c}\rangle\langle E_{n}^{c}|$. 
During this process only work is performed, with no heat being exchanged. 

3. Quantum isochoric process. The working substance
with frequency $\omega=\omega_{c}$ and modelled by the Hamiltonian $H_{Mf,c}^{s}$ is 
now put into contact with a cold reservoir at temperature $T_{c}<T_{h}$ until they reach thermal equilibrium. 
In this case, we have a change in the steady state population from $P_{n}^{ss}$($T_{h}$) to $P_{n}^{ss}$($T_{c}$), 
while the eigenvalues $E_{n}^{c}$ of the system remain unchanged, and the state 
becomes $\rho_{3}=\sum_{n}P_{n}^{ss}(T_{c})|E_{n}^{c}\rangle\langle E_{n}^{c}|$. 
During this process, only heat is exchanged, heat $Q_{c}$ is released to the cold reservoir, but no work is done. 

4. Quantum adiabatic compression process. The system is isolated
from the cold reservoir and its energy levels is changed back from
$E_{n}^{c}$ to $E_{n}^{h}$ by varying the frequency from $\omega_{c}$ to
$\omega_{h}$. At the end of the process, the populations $P_{n}^{ss}$
($T_{c}$) remain unchanged, the state becomes $\rho_{4}=\sum_{n}P_{n}^{ss}(T_{c})|E_{n}^{h}\rangle\langle E_{n}^{h}|$, 
and only work is performed by the working substance, but no heat is exchanged. 

Next, let us calculate the work and heat exchanged in each stroke. According to the first law of thermodynamics, 
a quantum system with discrete energy levels can be written as 
\begin{eqnarray}
dU=\delta Q+\delta W=\sum_{n}(E_{n}dP_{n}^{ss}+P_{n}^{ss}dE_{n}),
\end{eqnarray}
where $E_{n}$ are the energy levels and $P_{n}^{ss}$ are the populations at steady state.
Accordingly, the heat $Q_{h}$ ($Q_{c}$) exchanges with the hot (cold) reservoir, and the net 
work $W$ satisfy the following  relations: \cite{kieu2004second}
\begin{eqnarray}
Q_{h}=\sum_{n}E_{n}^{h}[P_{n}^{ss}(T_{h})-P_{n}^{ss}(T_{c})],
\end{eqnarray}
\begin{eqnarray}
Q_{c}=\sum_{n}E_{n}^{c}[P_{n}^{ss}(T_{c})-P_{n}^{ss}(T_{h})],
\end{eqnarray}
\begin{eqnarray}
W&=&Q_{h}+Q_{c}\nonumber\\
&=&\sum_{n}(E_{n}^{h}-E_{n}^{c})[P_{n}^{ss}(T_{h})-P_{n}^{ss}(T_{c})].
\end{eqnarray}

In this work we will adopt the following convention: $Q>0$ ($Q<0$) correspond to absorption (release) of heat from (to)
the reservoir while $W>0$ ($W<0$) correspond to work performed by
(on) the quantum heat cycle. There are only four working regimes
allowed under not violating the Clausius inequality with the first
law of thermodynamics \cite{solfanelli2020nonadiabatic}: 

\begin{table}[htb]   
\begin{center}   
\caption{Classification of working modes}  
\label{table:1} 
\begin{tabular}{|c|c|c|c|}   
\hline   \textbf{working regimes} & \textbf{$Q_h$} & \textbf{$Q_c$} & \textbf{$W$} \\   
\hline   Heat engine (E) & $>0$ & $<0$ & $>0$  \\ 
\hline   Refrigerator (R) & $<0$ & $>0$ & $<0$  \\  
\hline   Heater (H) & $<0$ & $<0$ & $<0$  \\  
\hline   Accelerator (A) & $>0$ & $>0$ & $<0$  \\ 
\hline   
\end{tabular}   
\end{center}   
\end{table}

In the following, we are more concerned with the
heat engine, which are of most interest for useful applications and whose figures of merit are the efficiency
$\eta=\frac{W}{Q_{h}}$.

%==========================================
\begin{figure}[htp]
%\begin{center}
%\vspace{-2.2cm}
\includegraphics[width=0.5\textwidth]{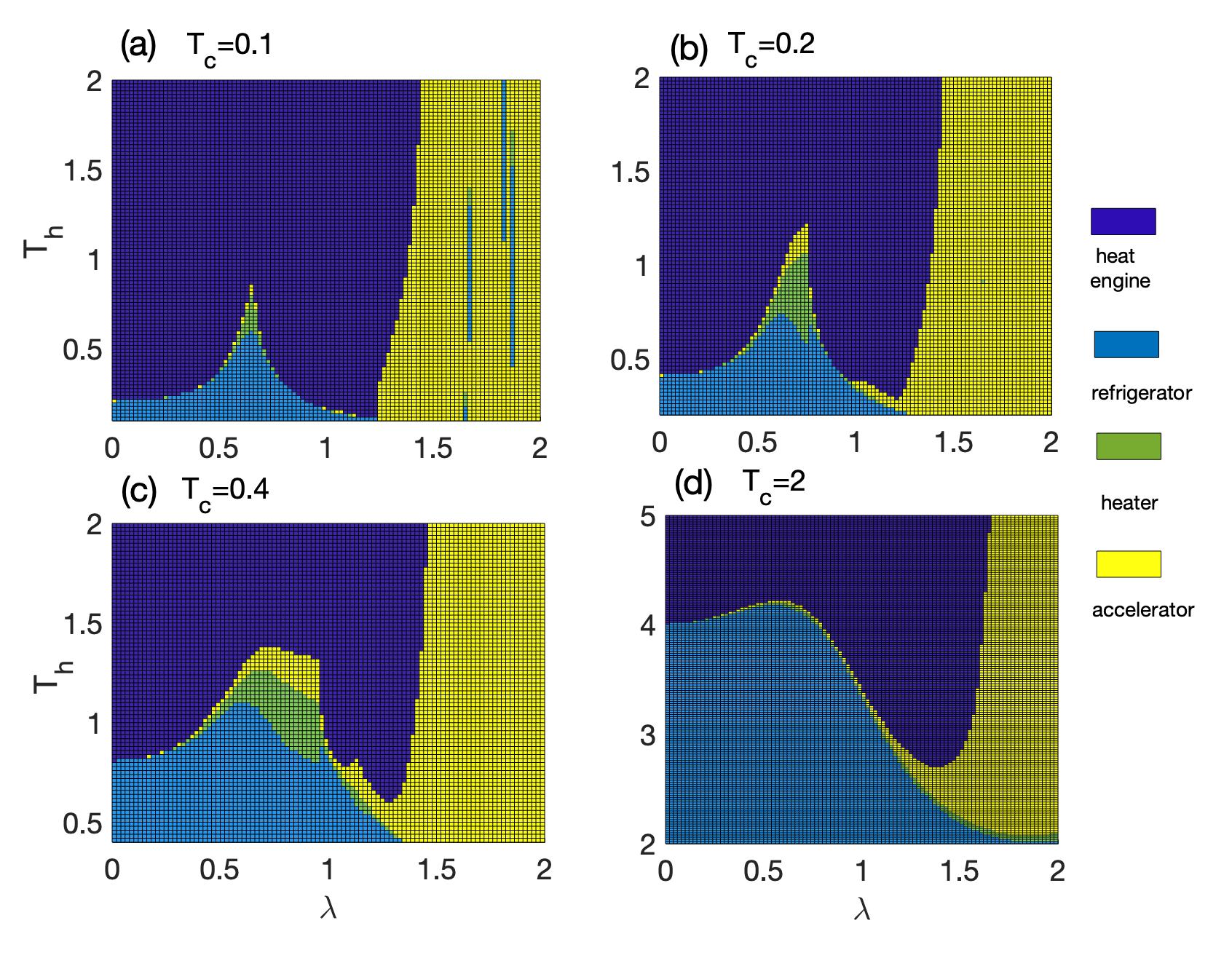}
%\vspace{-1.0cm}
%\end{center}
\caption{Phase diagrams of the working modes of quantum Otto cycle in the $T_{h}$-$\lambda$ parameter space under different $T_{c}$. 
 (a) $T_{c}=0.1$  (b) $T_{c}=0.2$, (c) $T_{c}=0.4$, and (d) $T_{c}=2$. The color code stands for heat engine (dark blue), refrigerator (light blue), 
 heater (grass green), and accelerator (yellow).
The other involved parameters are $N=8$, $J=0.01$, $\omega_{h} = 2\omega$, $\omega_{c} = \omega$, $\gamma_{c}=\gamma_{q}=10^{-4}$.}~\label{HdickeW}
\end{figure}
%========================================== 

%\begin{figure}[H]
%\vspace{-2.2cm}
%\includegraphics[width=0.5\textwidth]{Csmean11.eps}
%\vspace{-1.0cm}
%\caption{(a) The superfluid order parameter $|\psi|$ as a function of qubit-boson coupling strength $\lambda$ under 
%different $N$. 
%(b) The superfluid order parameter $|\psi|$ as a function of qubit-boson coupling strength $\lambda$ under 
%different inter-cavity hopping strength $J$. 
%}~\label{Csmean11}
%\end{figure}

\section{Results and Discussions}\label{S4}

Firstly, we study the influence of different heat source temperatures and atom-light coupling strengths on the working mode of 
the quantum Otto cycle. Figure \ref{HdickeW} shows the variation of the working mode of the quantum Otto heat engine with 
the temperature $T_{h}$ of the high-temperature heat source and the atom-light coupling strength $\lambda$ under 
different low-temperature heat source temperatures $T_{c}$. Different colors represent different working modes: dark blue is the 
heat engine mode, light blue is the refrigerator mode, grass green is the heater mode, and yellow is the accelerator mode.  

From Fig.~\ref{HdickeW}(a), we can see that the heat engine mode occupies most of the area with a relatively small $\lambda$, 
indicating that when $T_{c}$ is extremely low, the heat engine mode is easy to implement under weak coupling condition. 
The working area of the refrigeration machine mode is narrow, existing only when $\lambda$ is small and $T_h$ is within a 
specific range, and its application scope is limited. As $\lambda$ increases, the system quickly enters the accelerator mode, 
indicating that under strong coupling, the Otto cycle is more inclined to the accelerator function. Compared with the result of $T_{c}=0.1$, 
the mode distribution of $T_{c}=0.2$ has changed significantly. As shown in Fig.~\ref{HdickeW}(b), the mode range of the 
refrigerator mode has been significantly expanded, performing outstandingly within the range of 
moderate $\lambda$ and moderate $T_h$. Moreover, the area of the heater mode get larger. 
The heat engine mode still dominates in the smaller area of $\lambda$, but its relative proportion has slightly declined. The range 
of the accelerator mode in the larger area of $\lambda$ is wider than the result in Fig.~\ref{HdickeW}(a). 
For a slightly larger $T_{c}$ case, it is seen from Fig.~\ref{HdickeW}(c) that the 
 the working areas of the refrigerator, heater and accelerator modes further expand, 
while the working area of the heat engine is further compressed.
For higher $T_{c}$, It can be seen from Fig.~\ref{HdickeW}(d) that the area of the refrigerator mode has significantly 
expanded, becoming one of the core modes at this temperature.  The range of the heat engine mode has narrowed, 
concentrating in the regions with  higher $T_h$, and 
it almost disappears under strong coupling as well.  Although the heat engine, heater and accelerator modes still exist, 
the working area is significantly smaller compared to the results of $T_{c}=0.2$ and $T_{c}=0.4$, indicating that the 
higher $T_{c}$ can suppress the three modes.

%==========================================
\begin{figure}[htp]
%\begin{center}
%\vspace{-2.2cm}
\includegraphics[width=0.5\textwidth]{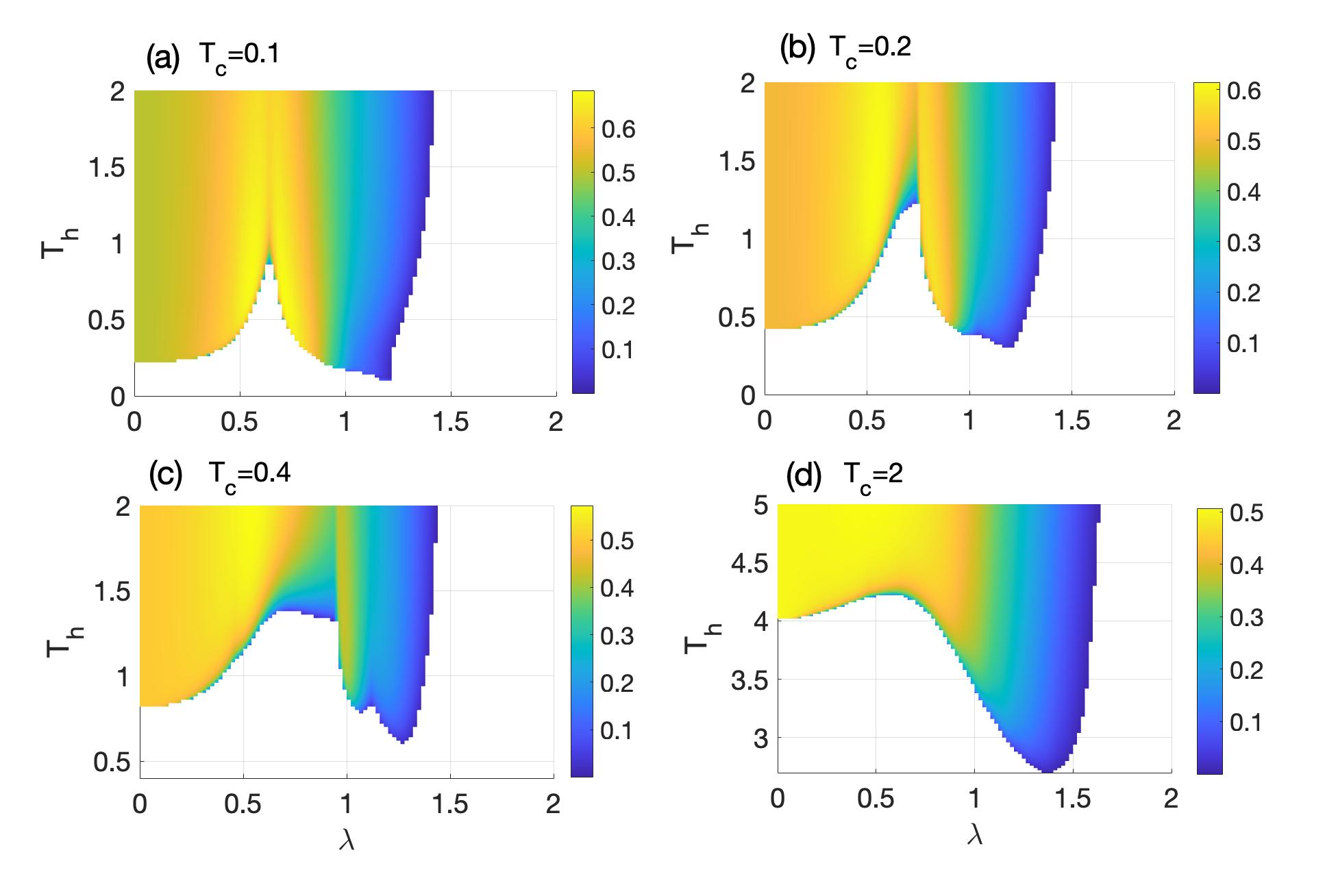}
%\vspace{-1.0cm}
%\end{center}
\caption{(a)-(c): The efficiency of the quantum Otto heat engine. (a) $T_{c}=0.1$;  (b) $T_{c}=0.2$; (c) $T_{c}=0.4$; (d) $T_{c}=2$. 
Other involved parameters are $N=8$, $J=0.01$, $\omega_{h} = 2\omega$, $\omega_{c} = \omega$, and $\gamma_{c}=\gamma_{q}=10^{-4}$.}~\label{Hdickeeta}
\end{figure}
%========================================== 

For different $T_{c}$, apart from causing differences in the distribution of working modes, they also bring 
some commonalities. For instance, the region $\lambda$ with a relatively high coupling strength is mainly in 
the accelerator mode, indicating that strong coupling is the core trigger condition of the accelerator mode and has nothing 
to do with $T_c$. Moreover, the heat engine mode is always distributed within a relatively small $\lambda$ and high $T_{h}$, 
indicating that relatively weak coupling and high temperature heat source is a fundamental condition of the heat engine mode. 
The refrigerator mode is always distributed within a relatively small $\lambda$ and low high $T_{h}$, 
indicating that relatively weak coupling and lower heat source temperature is a fundamental condition of the refrigerator mode.
These results provide a reference for regulating the working mode of the quantum Otto cycle.

Next, we analyze the  efficiency of  the quantum Otto heat engine. Figure \ref{Hdickeeta} shows the working efficiency of 
the quantum Otto heat engine in the $T_{h}$-$\lambda$ parameter space under different $T_c$ values. 
The color changes from yellow to green and then to blue, indicating that the work efficiency decreases from high to low. 
Figure \ref{Hdickeeta}(a) shows the result of $T_{c}=0.1$. It can be seen that the high work efficiency range covers a large 
area of $T_h \approx 0.5 \sim 2$ and $\lambda<1$, and the efficiency is the highest at moderate $\lambda$. When $\lambda>1$, 
the efficiency drops significantly, indicating the strong atom-light coupling is not conducive to the operation of heat engine. 
Figure \ref{Hdickeeta}(b) shows the result of $T_c=0.2$. Similar to the result of $T_c=0.1$, the high efficiency is mainly concentrated 
in the interval where $\lambda<1$, and the efficiency is the highest when the value of $\lambda$ is moderate. When the 
$J$ gets stronger ($\lambda>1$), the efficiency declines, signaling again that the strong coupling is not conducive 
to the operation of heat engine. For the $T_{c}=0.4$ case, it is seen from Fig.~\ref{Hdickeeta}(c) that the efficiency 
distribution characteristics are similar to those of $T_{c}=0.1$ and $T_{c}=0.2$. The high efficiency appears in the relatively weak 
coupling region, and the strong coupling is not conducive to the realization of the high-efficiency heat engine. 
The result of $T_c=2$ is presented in Fig.~\ref{Hdickeeta}(d). It can be seen that the bright 
yellow interval moves up to $T_h \approx 4\sim 5$, but it is also distributed in the weak coupling region ($\lambda<1$). When 
$\lambda>1$, efficiency drops rapidly as well. Through the study of heat engine efficiency, it has been clarified that weak 
coupling ($\lambda<1$) is the key to high efficiency, while strong coupling will lead to a decrease in efficiency. This result has 
guiding significance for the design of high-efficiency quantum Otto heat engines.
%==========================================
\begin{figure}[htp]
%\begin{center}
%\vspace{-2.2cm}
\includegraphics[width=0.5\textwidth]{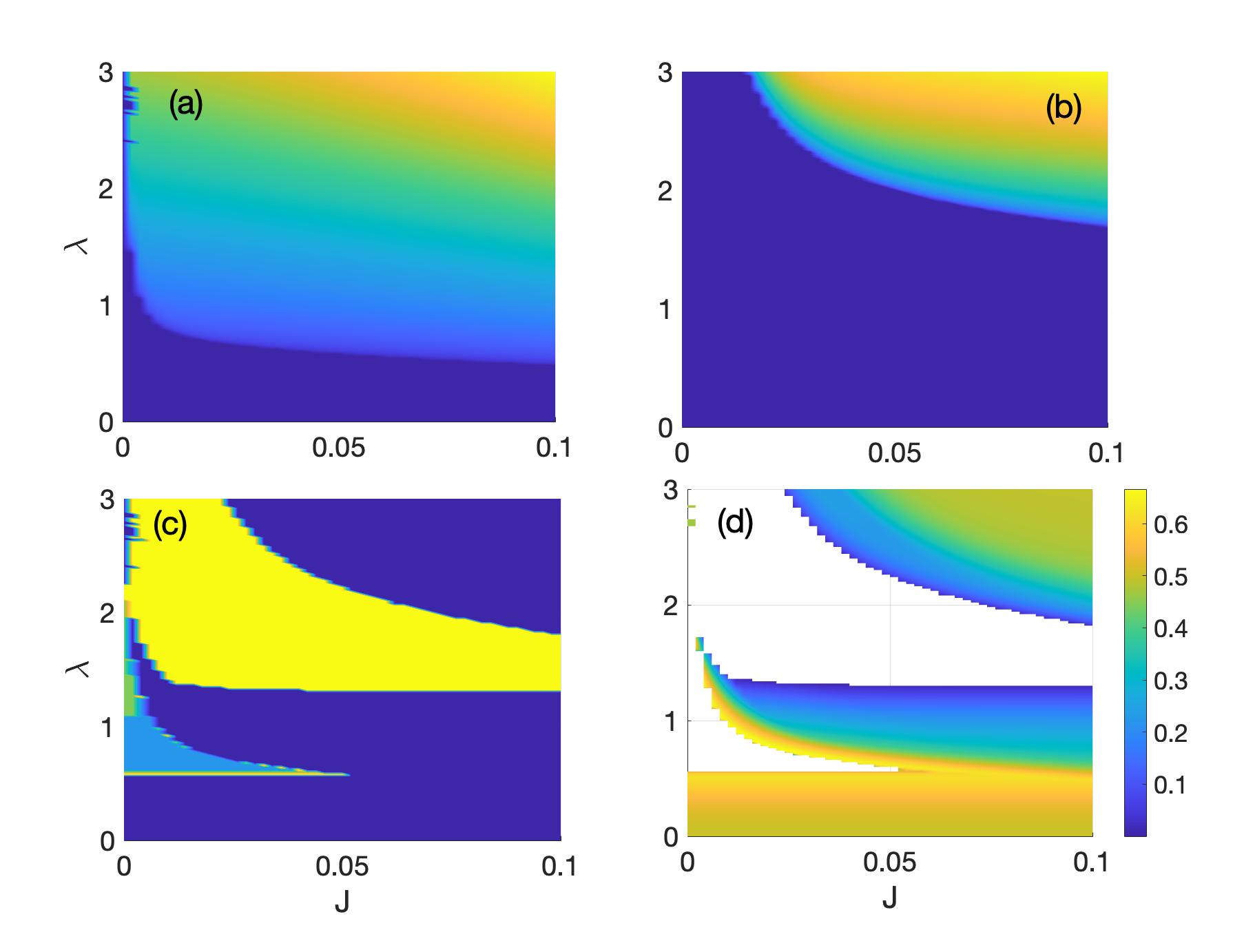}
%\vspace{-1.0cm}
%\end{center}
\caption{(a)-(b)The steady-state order parameter $|\psi|$ in $\lambda$-$J$ parameter space under temperature 
$T=0.1$ and $T=0.4$. 
(c) Phase diagrams of the working modes of quantum Otto cycle in the $\lambda$-$J$ parameter space 
with $T_c=0.1$ and $T_h=0.4$. 
(d) The efficiency of the quantum Otto heat engine. The other parameters are $N=4$, $\omega_{h} = 2\omega$, $\omega_{c} = \omega$, 
and $\gamma_{c}=\gamma_{q}=10^{-4}$.}~\label{HdickeLaJ}
\end{figure}
%========================================== 

In addition to the working mode of the quantum Otto cycle, another issue we are concerned about is the superradiance
phase transition. At zero temperature, the steady state of the system will undergo a superradiatance phase transition. Such phase transition 
can also occur at finite temperatures. Figures \ref{HdickeLaJ}(a)-\ref{HdickeLaJ}(b) show the steady state $|\psi|$ as functions 
of atom-light coupling strength $\lambda$ and inter-cavity hopping strength under temperatures $T=0.1$ and $T=0.4$, respectively. 
the dark blue regions denote the normal phase while others denote the superradiance phase. When $T=0.1$, it is seen that the 
superradiance phase covers a wide range of parameters which indicates that the system is prone to superradiance phase transition 
at low temperatures. For the $T=0.4$ case, the normal phase significantly expands, 
and only in the region with large $\lambda$ and moderate $J$ does the superradiance phase exist.

By comparing the results in Figs \ref{HdickeLaJ}(a) and \ref{HdickeLaJ} (b), we can draw a common rule: when $\lambda$is relatively 
small, regardless of whether the system is exposed to a low-temperature heat source or a high-temperature heat source, 
the steady state remains in the normal phase. When $\lambda$is relatively large, regardless of whether the system is exposed 
to a low-temperature heat source or a high-temperature heat source, the steady state remains in the normal phase. If, in an Otto cycle, 
the steady states of the systems in contact with high and low temperature heat sources are all in the same quantum phase, 
would this synergistic mechanism be conducive to the realization of the heat engine? To answer this question, we take $T_{c}=0.1$ 
and $T_{h}=0.4$ to calculate the working mode of the quantum Otto cycle, and the results are presented in Fig.~\ref{HdickeLaJ}(c). 
It can be seen that the working area of the heat engine completely covers the regions of the normal phase and the superradiance 
phase. This indicates that the synergistic effect of the low-temperature heat source and the high-temperature heat source helps the 
Otto cycle to achieve the heat engine function. In addition, it can be observed that the competition effect between low-temperature 
heat sources and high-temperature heat sources, i.e., the different steady-state properties at high and low temperatures, will give 
rise to a rich variety of working modes, including the refrigerator, heater and the accelerator, as well as the heat engine. We have 
known that both the synergistic effect and the competitive effect can trigger the heat engine. Then, under which mechanism is the 
working efficiency of heat engine higher? Regarding this issue, we calculate the efficiency of the heat engine, and the results are 
presented in Fig.~\ref{HdickeLaJ}(d). It can be seen that the high efficiency is mainly distributed in the area under the synergistic 
mechanism. In particular, when the steady states at both low and high temperatures remain in the normal phase, the efficiency is higher. 
This result tells us that if one wants the heat engine to work at high efficiency, preparing the working substance in the normal phase 
is a much better approach.

%==========================================
\begin{figure}[htp]
%\begin{center}
%\vspace{-2.2cm}
\includegraphics[width=0.5\textwidth]{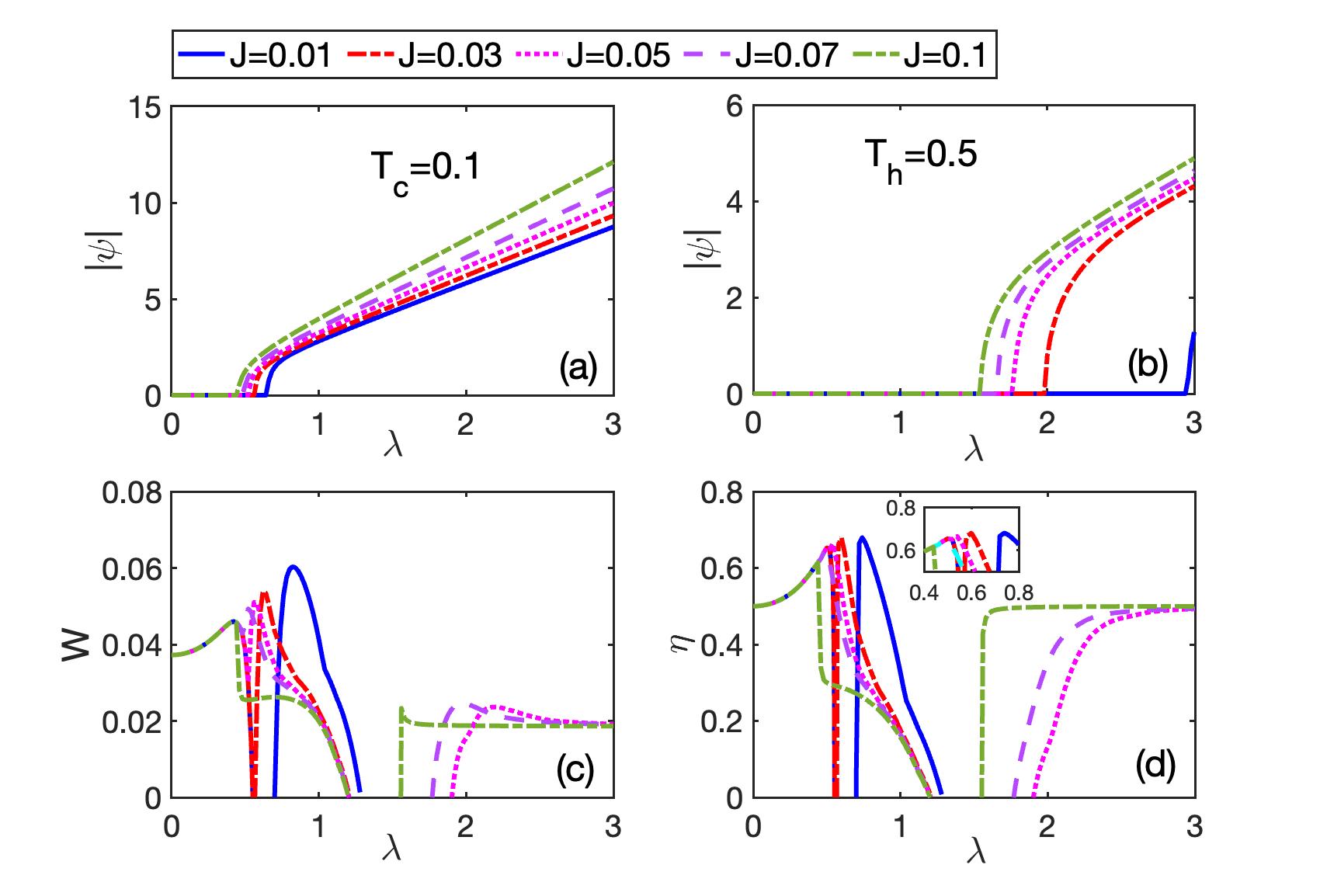}
%\vspace{-1.0cm}
%\end{center}
\caption{(a)-(b) The steady-state order parameter $|\psi|$ as the function of atom-light coupling strength $\lambda$ under 
different hopping strength $J$. The temperature is taken as (a) $T=0.1$, (b)$T=0.5$. 
(c)-(d) The efficiency $\eta$ and Work $W$ of the quantum heat engine as a function of the atom-light 
coupling strength $\lambda$ for different hopping strength $J$, keeping $T_c=0.1$ and $T_h=0.5$. 
The other system parameters are given 
by $N=8$, $\omega_{h} = 2\omega$, $\omega_{c} = \omega$, and $\gamma_{c}=\gamma_{q}=10^{-4}$.}~\label{HdickeJ}
\end{figure}
%========================================== 

In addition, we investigate the influence of different inter-cavity hopping strengths on the superradiance phase transition at different 
temperatures, and compare the differences in output work and efficiency under different $J$. The results are presented in 
Figs.~\ref{HdickeJ}(a)-\ref{HdickeJ}(d) respectively. As can be seen from Fig.~\ref{HdickeJ}(a), the larger $J$, the smaller 
the critical $\lambda$, indicates that the inter-cavity hopping strength $J$ can promote superradiance at 
low temperature.  However, the variation trend of the phase transition point is not very obvious as $J$ further increases.  Compared 
with the results in \ref{HdickeJ}(a), the superradiance phase transition point at high temperatures also shows a similar trend of 
change, but the difference lies in that the superradiance phase transition point is more sensitive to the change of $J$.  At the 
same time, it can be seen that at high temperatures, the non-zero regions of the order parameter $|\psi|$ corresponding to 
different $J$ significantly decrease, further verifying the suppression effect of temperature on the superradiance phase transition. 
 As can be seen from Fig.\ref{HdickeJ}(c), under the weak inter-cavity hopping condition, such as $J=0.01$, the heat engine has 
 the maximum output work, while under the larger, such as $J=0.1$, the peak output work is the lowest.  However, if it is necessary 
 for the heat engine to have good output power, it is essential to adjust $\lambda$ and $J$ within an appropriate range. Concretely, 
 fixing $\lambda$ in the weak atom-light coupling interval is a better choice. Figure \ref{HdickeJ}(d) reflects the working efficiency 
 of the heat engine, and the inset magnifies the details of the high-efficiency range.  It can be seen that the peak of efficiency 
 is not sensitive to the change of $J$.  Adjusting $\lambda$ and $J$ within an appropriate range  can enable the heat engine 
 to operate with good performance. Concretely, to ensure that the output work of the heat engine is no less than $0.04$, 
 it is a better choice to set the $\lambda$ in the weak atom-light coupling region.  When $\lambda$ is set in the appropriate 
 weak or strong atom-light coupling regions, the heat engine can achieve at least a $0.5$ working efficiency.

%==========================================
\begin{figure}[htp]
%\begin{center}
%\vspace{-2.2cm}
\includegraphics[width=0.5\textwidth]{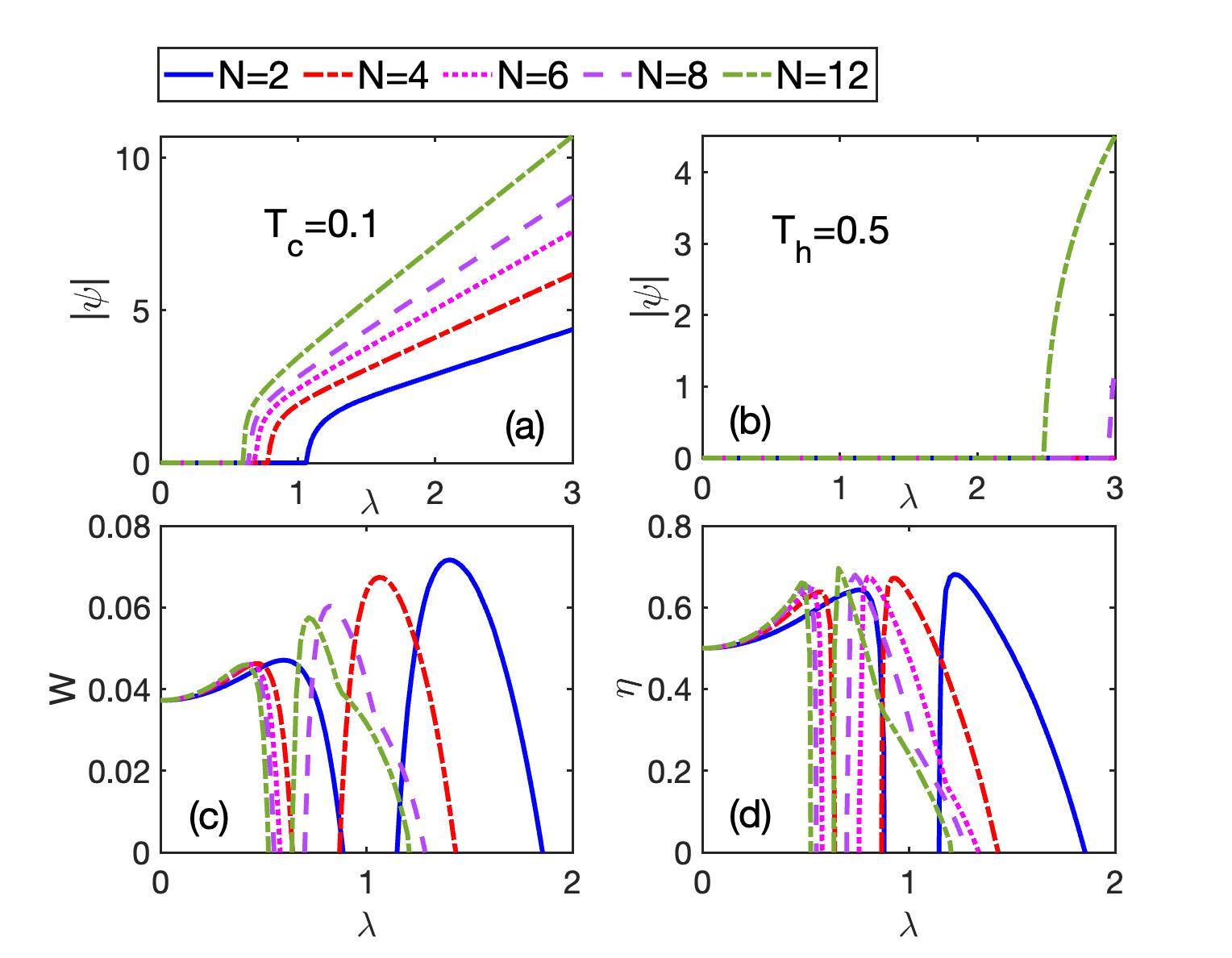}
%\vspace{-1.0cm}
%\end{center}
\caption{(a)-(b) The steady-state order parameter $|\psi|$ as the function of qubit-boson coupling 
strength $\lambda$ under different qubits number with fix hopping strength $J=0.01$. The temperature is 
taken as (a) $T=0.1$ and (b) $T=0.5$. (c)-(d) The efficiency $\eta$  and work $W$ of the quantum heat 
engine as a function of the qubit-boson coupling strength $\lambda$ for different qubit number with fix 
hopping strength $J=0.01$, $T_c=0.1$, and  $T_h=0.5$. The other parameters are $\omega_{h} = 2\omega$, 
$\omega_{c} = \omega$, and $\gamma_{c}=\gamma_{q}=10^{-4}$.}~\label{HdickeN}
\end{figure}
%==========================================

We also investigate the influence of different atomic numbers $N$ on the superradiance phase transition, output work 
and efficiency, and the results are presented in Figs.~\ref{HdickeN}(a) and \ref{HdickeN}(d) respectively.  As can be seen from 
Figs.~\ref{HdickeN}(a) and \ref{HdickeN}(b), the more $N$ is, the smaller the critical $\lambda$ of the phase transition becomes, 
indicating that a greater number of atoms can promote superradiance phase transitions.  However, through comparison, it is found 
that the changing trend of the phase transition critical point at high temperatures is more obvious than that at low temperatures. 
As can be seen from Fig.~\ref{HdickeN}(b), the superradiance phase transition will only occur when the number of atoms is large 
and the atom-light coupling strength is very strong.  Here, the suppression effect of temperature on the superradiative phase 
transition is verified.  As can be seen from Fig.~\ref{HdickeN}(c), the output work of the heat engine depends on the coordinated 
regulation of $\lambda$ and $N$.  When $N$ is smaller, the peak value of the output work is higher.  Besides, we can see that regardless 
of whether the atom-light coupling strength is weak or strong, the appropriate atomic number conditions can always be found to 
keep the output work no less than $0.04$. The phenomena are slightly different from those under different $J$ 
conditions.  In  Fig.~\ref{HdickeN}(d), we can see that no matter it is the strong atom-light coupling or the weak coupling, 
the appropriate atomic number conditions can always be found, which enables the heat engine to have a good performance 
with a working efficiency no less than $0.5$.

%==========================================
%\begin{figure}[tbp]
%\begin{center}
%\vspace{-2.2cm}
%\includegraphics[width=0.5\textwidth]{HdickeG2.jpg}
%\vspace{-1.0cm}
%\end{center}
%\caption{(a)-(b)Two-photon correlation function $G^{(2)}_8(0)$ as a function of qubit-photon coupling strength $\lambda$ under different hopping strength $J$ with fix qubits number $N=8$. The temperature is taken as (a)$T_c=0.1$, (b)$T_h=0.5$.
 %(c)-(d)Two-photon correlation function $G^{(2)}_8(0)$ as a function of qubit-photon coupling strength $\lambda$ under different qubits number with hopping strength $J=0.01$. The temperature is taken as (c)$T_c=0.1$, (d)$T_h=0.5$. The other system parameters are given by $\omega_{h} = 2\omega$, $\omega_{c} = \omega$, $\gamma_{c}=\gamma_{q}=10^{-4}$.}~\label{HdickeG2}
%\end{figure}
%========================================== 

\section{Summary}\label{S5}
In summary, this work has studied a quantum Otto heat engine whose working substance is governed by the dissipative 
Dicke-Hubbard model. By employing a mean-field approximation, extended bosonic coherent state approach and a 
quantum dressed master equation approach, we have self-consistently determined the properties of system's steady states in contact 
with thermal baths at different temperatures. Our investigation reveals how the engine's operational mode—heat engine, refrigerator, heater, or accelerator—is controlled by key parameters, including the atom-light coupling strength ($\lambda$), inter-cavity hopping ($J$), number of atoms ($N$), and bath temperatures ($T_h$,$T_c$). A central finding is that high-performance heat engine operation, characterized by high efficiency, predominantly occurs 
in the relatively weak atom-light coupling regime ($\lambda<1$). Strong coupling generally drives the system into an accelerator mode 
and diminishes engine efficiency. Furthermore, we have established a direct connection between the engine's performance and the 
underlying superradiance phase transition of the DH model. We identify that a synergistic mechanism, where the system remains in 
the same quantum phase (particularly the normal phase) at both high and low temperatures, is highly conducive to achieving high 
work efficiency. In contrast, a competitive mechanism, where the steady states at the two temperatures belong to different phases, 
gives rise to a richer variety of operational modes but typically yields lower efficiency. This work deepens the understanding of energy 
conversion in non-equilibrium quantum many-body systems and provides concrete theoretical guidance for the design of 
high-performance quantum heat machines by leveraging collective quantum phenomena.

\begin{acknowledgments}
 This work is supported by the Natural Science Foundation of Zhejiang Province (Grants No. LQN25A040012) and
the start-up fund from Xingzhi college, Zhejiang Normal University.
\end{acknowledgments}

\bibliography{references}

\begin{thebibliography}{77}%
\makeatletter
\providecommand \@ifxundefined [1]{%
 \@ifx{#1\undefined}
}%
\providecommand \@ifnum [1]{%
 \ifnum #1\expandafter \@firstoftwo
 \else \expandafter \@secondoftwo
 \fi
}%
\providecommand \@ifx [1]{%
 \ifx #1\expandafter \@firstoftwo
 \else \expandafter \@secondoftwo
 \fi
}%
\providecommand \natexlab [1]{#1}%
\providecommand \enquote  [1]{``#1''}%
\providecommand \bibnamefont  [1]{#1}%
\providecommand \bibfnamefont [1]{#1}%
\providecommand \citenamefont [1]{#1}%
\providecommand \href@noop [0]{\@secondoftwo}%
\providecommand \href [0]{\begingroup \@sanitize@url \@href}%
\providecommand \@href[1]{\@@startlink{#1}\@@href}%
\providecommand \@@href[1]{\endgroup#1\@@endlink}%
\providecommand \@sanitize@url [0]{\catcode `\\12\catcode `\$12\catcode
  `\&12\catcode `\#12\catcode `\^12\catcode `\_12\catcode `\%12\relax}%
\providecommand \@@startlink[1]{}%
\providecommand \@@endlink[0]{}%
\providecommand \url  [0]{\begingroup\@sanitize@url \@url }%
\providecommand \@url [1]{\endgroup\@href {#1}{\urlprefix }}%
\providecommand \urlprefix  [0]{URL }%
\providecommand \Eprint [0]{\href }%
\providecommand \doibase [0]{http://dx.doi.org/}%
\providecommand \selectlanguage [0]{\@gobble}%
\providecommand \bibinfo  [0]{\@secondoftwo}%
\providecommand \bibfield  [0]{\@secondoftwo}%
\providecommand \translation [1]{[#1]}%
\providecommand \BibitemOpen [0]{}%
\providecommand \bibitemStop [0]{}%
\providecommand \bibitemNoStop [0]{.\EOS\space}%
\providecommand \EOS [0]{\spacefactor3000\relax}%
\providecommand \BibitemShut  [1]{\csname bibitem#1\endcsname}%
\let\auto@bib@innerbib\@empty
%</preamble>
\bibitem [{\citenamefont {Vinjanampathy}\ and\ \citenamefont
  {Anders}(2016)}]{Vinjanampathy2016}%
  \BibitemOpen
  \bibfield  {author} {\bibinfo {author} {\bibfnamefont {S.}~\bibnamefont
  {Vinjanampathy}}\ and\ \bibinfo {author} {\bibfnamefont {J.}~\bibnamefont
  {Anders}},\ }\bibfield  {title} {\enquote {\bibinfo {title} {Quantum
  thermodynamics},}\ }\href {\doibase 10.1080/00107514.2016.1201896} {\bibfield
   {journal} {\bibinfo  {journal} {Contemporary Physics}\ }\textbf {\bibinfo
  {volume} {57}},\ \bibinfo {pages} {545--579} (\bibinfo {year}
  {2016})}\BibitemShut {NoStop}%
\bibitem [{\citenamefont {Millen}\ and\ \citenamefont
  {Xuereb}(2016)}]{Millen2016}%
  \BibitemOpen
  \bibfield  {author} {\bibinfo {author} {\bibfnamefont {J.}~\bibnamefont
  {Millen}}\ and\ \bibinfo {author} {\bibfnamefont {A.}~\bibnamefont
  {Xuereb}},\ }\bibfield  {title} {\enquote {\bibinfo {title} {Perspective on
  quantum thermodynamics},}\ }\href {\doibase 10.1088/1367-2630/18/1/011002}
  {\bibfield  {journal} {\bibinfo  {journal} {New Journal of Physics}\ }\textbf
  {\bibinfo {volume} {18}},\ \bibinfo {pages} {011002} (\bibinfo {year}
  {2016})}\BibitemShut {NoStop}%
\bibitem [{\citenamefont {Kosloff}(2013)}]{Kosloff2013}%
  \BibitemOpen
  \bibfield  {author} {\bibinfo {author} {\bibfnamefont {R.}~\bibnamefont
  {Kosloff}},\ }\bibfield  {title} {\enquote {\bibinfo {title} {Quantum
  thermodynamics: A dynamical viewpoint},}\ }\href {\doibase 10.3390/e15062100}
  {\bibfield  {journal} {\bibinfo  {journal} {Entropy}\ }\textbf {\bibinfo
  {volume} {15}},\ \bibinfo {pages} {2100--2128} (\bibinfo {year}
  {2013})}\BibitemShut {NoStop}%
\bibitem [{\citenamefont {Quan}\ \emph
  {et~al.}(2007{\natexlab{a}})\citenamefont {Quan}, \citenamefont {Liu},
  \citenamefont {Sun},\ and\ \citenamefont {Nori}}]{Quan2007}%
  \BibitemOpen
  \bibfield  {author} {\bibinfo {author} {\bibfnamefont {H.~T.}\ \bibnamefont
  {Quan}}, \bibinfo {author} {\bibfnamefont {Y.~X.}\ \bibnamefont {Liu}},
  \bibinfo {author} {\bibfnamefont {C.~P.}\ \bibnamefont {Sun}}, \ and\
  \bibinfo {author} {\bibfnamefont {F.}~\bibnamefont {Nori}},\ }\bibfield
  {title} {\enquote {\bibinfo {title} {Quantum thermodynamic cycles and quantum
  heat engines},}\ }\href {\doibase 10.1103/PhysRevE.76.031105} {\bibfield
  {journal} {\bibinfo  {journal} {Physical Review E}\ }\textbf {\bibinfo
  {volume} {76}},\ \bibinfo {pages} {031105} (\bibinfo {year}
  {2007}{\natexlab{a}})}\BibitemShut {NoStop}%
\bibitem [{\citenamefont {Uzdin}\ \emph {et~al.}(2015)\citenamefont {Uzdin},
  \citenamefont {Levy},\ and\ \citenamefont {Kosloff}}]{Uzdin2015}%
  \BibitemOpen
  \bibfield  {author} {\bibinfo {author} {\bibfnamefont {R.}~\bibnamefont
  {Uzdin}}, \bibinfo {author} {\bibfnamefont {A.}~\bibnamefont {Levy}}, \ and\
  \bibinfo {author} {\bibfnamefont {R.}~\bibnamefont {Kosloff}},\ }\bibfield
  {title} {\enquote {\bibinfo {title} {Equivalence of quantum heat machines,
  and quantum-thermodynamic signatures},}\ }\href {\doibase
  10.1103/PhysRevX.5.031044} {\bibfield  {journal} {\bibinfo  {journal}
  {Physical Review X}\ }\textbf {\bibinfo {volume} {5}},\ \bibinfo {pages}
  {031044} (\bibinfo {year} {2015})}\BibitemShut {NoStop}%
\bibitem [{\citenamefont {Campisi}\ and\ \citenamefont
  {Fazio}(2016)}]{Campisi2016}%
  \BibitemOpen
  \bibfield  {author} {\bibinfo {author} {\bibfnamefont {M.}~\bibnamefont
  {Campisi}}\ and\ \bibinfo {author} {\bibfnamefont {R.}~\bibnamefont
  {Fazio}},\ }\bibfield  {title} {\enquote {\bibinfo {title} {The power of a
  critical heat engine},}\ }\href {\doibase 10.1038/ncomms11895} {\bibfield
  {journal} {\bibinfo  {journal} {Nature Communications}\ }\textbf {\bibinfo
  {volume} {7}},\ \bibinfo {pages} {11895} (\bibinfo {year}
  {2016})}\BibitemShut {NoStop}%
\bibitem [{\citenamefont {Zhang}\ \emph
  {et~al.}(2014{\natexlab{a}})\citenamefont {Zhang}, \citenamefont {Bariani},\
  and\ \citenamefont {Dong}}]{Zhang2014}%
  \BibitemOpen
  \bibfield  {author} {\bibinfo {author} {\bibfnamefont {K.}~\bibnamefont
  {Zhang}}, \bibinfo {author} {\bibfnamefont {F.}~\bibnamefont {Bariani}}, \
  and\ \bibinfo {author} {\bibfnamefont {Y.}~\bibnamefont {Dong}},\ }\bibfield
  {title} {\enquote {\bibinfo {title} {Quantum optomechanical heat engine},}\
  }\href {\doibase 10.1103/PhysRevLett.112.150602} {\bibfield  {journal}
  {\bibinfo  {journal} {Physical Review Letters}\ }\textbf {\bibinfo {volume}
  {112}},\ \bibinfo {pages} {150602} (\bibinfo {year}
  {2014}{\natexlab{a}})}\BibitemShut {NoStop}%
\bibitem [{\citenamefont {Feldmann}\ and\ \citenamefont
  {Kosloff}(2000)}]{feldmann2000performance}%
  \BibitemOpen
  \bibfield  {author} {\bibinfo {author} {\bibfnamefont {T.}~\bibnamefont
  {Feldmann}}\ and\ \bibinfo {author} {\bibfnamefont {R.}~\bibnamefont
  {Kosloff}},\ }\bibfield  {title} {\enquote {\bibinfo {title} {Performance of
  discrete heat engines and heat pumps in finite time},}\ }\href {\doibase
  10.1103/PhysRevE.61.4774} {\bibfield  {journal} {\bibinfo  {journal}
  {Physical Review E}\ }\textbf {\bibinfo {volume} {61}},\ \bibinfo {pages}
  {4774} (\bibinfo {year} {2000})}\BibitemShut {NoStop}%
\bibitem [{\citenamefont {Geva}\ and\ \citenamefont
  {Kosloff}(1992)}]{geva1992quantum}%
  \BibitemOpen
  \bibfield  {author} {\bibinfo {author} {\bibfnamefont {E.}~\bibnamefont
  {Geva}}\ and\ \bibinfo {author} {\bibfnamefont {R.}~\bibnamefont {Kosloff}},\
  }\bibfield  {title} {\enquote {\bibinfo {title} {A quantum-mechanical heat
  engine operating in finite time. a model consisting of spin-1/2 systems as
  the working fluid},}\ }\href {\doibase 10.1063/1.461951} {\bibfield
  {journal} {\bibinfo  {journal} {The Journal of chemical physics}\ }\textbf
  {\bibinfo {volume} {96}},\ \bibinfo {pages} {3054--3067} (\bibinfo {year}
  {1992})}\BibitemShut {NoStop}%
\bibitem [{\citenamefont {de~Assis}\ \emph {et~al.}(2021)\citenamefont
  {de~Assis}, \citenamefont {Sales}, \citenamefont {Mendes},\ and\
  \citenamefont {de~Almeida}}]{deAssis_2021}%
  \BibitemOpen
  \bibfield  {author} {\bibinfo {author} {\bibfnamefont {R.~J.}\ \bibnamefont
  {de~Assis}}, \bibinfo {author} {\bibfnamefont {J.~S.}\ \bibnamefont {Sales}},
  \bibinfo {author} {\bibfnamefont {U.~C.}\ \bibnamefont {Mendes}}, \ and\
  \bibinfo {author} {\bibfnamefont {N.~G.}\ \bibnamefont {de~Almeida}},\
  }\bibfield  {title} {\enquote {\bibinfo {title} {Two-level quantum otto heat
  engine operating with unit efficiency far from the quasi-static regime under
  a squeezed reservoir},}\ }\href {\doibase 10.1088/1361-6455/abcfd9}
  {\bibfield  {journal} {\bibinfo  {journal} {Journal of Physics B: Atomic,
  Molecular and Optical Physics}\ }\textbf {\bibinfo {volume} {54}},\ \bibinfo
  {pages} {095501} (\bibinfo {year} {2021})}\BibitemShut {NoStop}%
\bibitem [{\citenamefont {Ro{\ss}nagel}\ \emph
  {et~al.}(2014{\natexlab{a}})\citenamefont {Ro{\ss}nagel}, \citenamefont
  {Abah}, \citenamefont {Schmidt-Kaler}, \citenamefont {Singer},\ and\
  \citenamefont {Lutz}}]{Rossnagel2014}%
  \BibitemOpen
  \bibfield  {author} {\bibinfo {author} {\bibfnamefont {J.}~\bibnamefont
  {Ro{\ss}nagel}}, \bibinfo {author} {\bibfnamefont {O.}~\bibnamefont {Abah}},
  \bibinfo {author} {\bibfnamefont {F.}~\bibnamefont {Schmidt-Kaler}}, \bibinfo
  {author} {\bibfnamefont {K.}~\bibnamefont {Singer}}, \ and\ \bibinfo {author}
  {\bibfnamefont {E.}~\bibnamefont {Lutz}},\ }\bibfield  {title} {\enquote
  {\bibinfo {title} {Nanoscale heat engine beyond the {Carnot} limit},}\ }\href
  {\doibase 10.1103/PhysRevLett.112.030602} {\bibfield  {journal} {\bibinfo
  {journal} {Physical Review Letters}\ }\textbf {\bibinfo {volume} {112}},\
  \bibinfo {pages} {030602} (\bibinfo {year} {2014}{\natexlab{a}})}\BibitemShut
  {NoStop}%
\bibitem [{\citenamefont {Abah}\ \emph {et~al.}(2012)\citenamefont {Abah},
  \citenamefont {Rossnagel}, \citenamefont {Jacob}, \citenamefont {Deffner},
  \citenamefont {Schmidt-Kaler}, \citenamefont {Singer},\ and\ \citenamefont
  {Lutz}}]{abah2012single}%
  \BibitemOpen
  \bibfield  {author} {\bibinfo {author} {\bibfnamefont {O.}~\bibnamefont
  {Abah}}, \bibinfo {author} {\bibfnamefont {J.}~\bibnamefont {Rossnagel}},
  \bibinfo {author} {\bibfnamefont {G.}~\bibnamefont {Jacob}}, \bibinfo
  {author} {\bibfnamefont {S.}~\bibnamefont {Deffner}}, \bibinfo {author}
  {\bibfnamefont {F.}~\bibnamefont {Schmidt-Kaler}}, \bibinfo {author}
  {\bibfnamefont {K.}~\bibnamefont {Singer}}, \ and\ \bibinfo {author}
  {\bibfnamefont {E.}~\bibnamefont {Lutz}},\ }\bibfield  {title} {\enquote
  {\bibinfo {title} {Single-ion heat engine at maximum power},}\ }\href
  {\doibase 10.1103/PhysRevLett.109.203006} {\bibfield  {journal} {\bibinfo
  {journal} {Physical review letters}\ }\textbf {\bibinfo {volume} {109}},\
  \bibinfo {pages} {203006} (\bibinfo {year} {2012})}\BibitemShut {NoStop}%
\bibitem [{\citenamefont {Myers}\ and\ \citenamefont
  {Deffner}(2020)}]{myers2020bosons}%
  \BibitemOpen
  \bibfield  {author} {\bibinfo {author} {\bibfnamefont {N.~M.}\ \bibnamefont
  {Myers}}\ and\ \bibinfo {author} {\bibfnamefont {S.}~\bibnamefont
  {Deffner}},\ }\bibfield  {title} {\enquote {\bibinfo {title} {Bosons
  outperform fermions: The thermodynamic advantage of symmetry},}\ }\href
  {\doibase 10.1103/PhysRevE.101.012110} {\bibfield  {journal} {\bibinfo
  {journal} {Physical Review E}\ }\textbf {\bibinfo {volume} {101}},\ \bibinfo
  {pages} {012110} (\bibinfo {year} {2020})}\BibitemShut {NoStop}%
\bibitem [{\citenamefont {Klaers}\ \emph {et~al.}(2017)\citenamefont {Klaers},
  \citenamefont {Faelt}, \citenamefont {Imamoglu},\ and\ \citenamefont
  {Togan}}]{Klaers2017}%
  \BibitemOpen
  \bibfield  {author} {\bibinfo {author} {\bibfnamefont {J.}~\bibnamefont
  {Klaers}}, \bibinfo {author} {\bibfnamefont {S.}~\bibnamefont {Faelt}},
  \bibinfo {author} {\bibfnamefont {A.}~\bibnamefont {Imamoglu}}, \ and\
  \bibinfo {author} {\bibfnamefont {E.}~\bibnamefont {Togan}},\ }\bibfield
  {title} {\enquote {\bibinfo {title} {Squeezed thermal reservoirs as a
  resource for a nanomechanical heat engine},}\ }\href {\doibase
  10.1103/PhysRevX.7.031044} {\bibfield  {journal} {\bibinfo  {journal}
  {Physical Review X}\ }\textbf {\bibinfo {volume} {7}},\ \bibinfo {pages}
  {031044} (\bibinfo {year} {2017})}\BibitemShut {NoStop}%
\bibitem [{\citenamefont {Wang}\ \emph {et~al.}(2012)\citenamefont {Wang},
  \citenamefont {Wu},\ and\ \citenamefont {He}}]{2012Quantum}%
  \BibitemOpen
  \bibfield  {author} {\bibinfo {author} {\bibfnamefont {J.}~\bibnamefont
  {Wang}}, \bibinfo {author} {\bibfnamefont {Z.}~\bibnamefont {Wu}}, \ and\
  \bibinfo {author} {\bibfnamefont {J.}~\bibnamefont {He}},\ }\bibfield
  {title} {\enquote {\bibinfo {title} {Quantum otto engine of a two-level atom
  with single-mode fields},}\ }\href {\doibase 10.1103/PhysRevE.85.041148}
  {\bibfield  {journal} {\bibinfo  {journal} {Phys Rev E Stat Nonlin Soft
  Matter Phys}\ }\textbf {\bibinfo {volume} {85}},\ \bibinfo {pages} {041148}
  (\bibinfo {year} {2012})}\BibitemShut {NoStop}%
\bibitem [{\citenamefont {Feldmann}\ and\ \citenamefont
  {Kosloff}(2004)}]{2004Characteristics}%
  \BibitemOpen
  \bibfield  {author} {\bibinfo {author} {\bibfnamefont {T.}~\bibnamefont
  {Feldmann}}\ and\ \bibinfo {author} {\bibfnamefont {R.}~\bibnamefont
  {Kosloff}},\ }\bibfield  {title} {\enquote {\bibinfo {title} {Characteristics
  of the limit cycle of a reciprocating quantum heat engine},}\ }\href
  {\doibase 10.1103/PhysRevE.70.046110} {\bibfield  {journal} {\bibinfo
  {journal} {Physical Review E}\ }\textbf {\bibinfo {volume} {70}},\ \bibinfo
  {pages} {046110} (\bibinfo {year} {2004})}\BibitemShut {NoStop}%
\bibitem [{\citenamefont {Huang}\ \emph {et~al.}(2013)\citenamefont {Huang},
  \citenamefont {Xu}, \citenamefont {Niu},\ and\ \citenamefont
  {Fu}}]{huang2013special}%
  \BibitemOpen
  \bibfield  {author} {\bibinfo {author} {\bibfnamefont {X.~L.}\ \bibnamefont
  {Huang}}, \bibinfo {author} {\bibfnamefont {H.}~\bibnamefont {Xu}}, \bibinfo
  {author} {\bibfnamefont {X.~Y.}\ \bibnamefont {Niu}}, \ and\ \bibinfo
  {author} {\bibfnamefont {Y.~D.}\ \bibnamefont {Fu}},\ }\bibfield  {title}
  {\enquote {\bibinfo {title} {A special entangled quantum heat engine based on
  the two-qubit heisenberg xx model},}\ }\href {\doibase
  10.1088/0031-8949/88/06/065008} {\bibfield  {journal} {\bibinfo  {journal}
  {Physica Scripta}\ }\textbf {\bibinfo {volume} {88}},\ \bibinfo {pages}
  {065008} (\bibinfo {year} {2013})}\BibitemShut {NoStop}%
\bibitem [{\citenamefont {Huang}\ \emph {et~al.}(2014)\citenamefont {Huang},
  \citenamefont {Liu}, \citenamefont {Wang},\ and\ \citenamefont
  {Niu}}]{huang2014specialf}%
  \BibitemOpen
  \bibfield  {author} {\bibinfo {author} {\bibfnamefont {X.~L.}\ \bibnamefont
  {Huang}}, \bibinfo {author} {\bibfnamefont {Y.}~\bibnamefont {Liu}}, \bibinfo
  {author} {\bibfnamefont {Z.}~\bibnamefont {Wang}}, \ and\ \bibinfo {author}
  {\bibfnamefont {X.~Y.}\ \bibnamefont {Niu}},\ }\bibfield  {title} {\enquote
  {\bibinfo {title} {Special coupled quantum otto cycles},}\ }\href {\doibase
  10.1140/epjp/i2014-14004-8} {\bibfield  {journal} {\bibinfo  {journal} {The
  European Physical Journal Plus}\ }\textbf {\bibinfo {volume} {129}},\
  \bibinfo {pages} {4} (\bibinfo {year} {2014})}\BibitemShut {NoStop}%
\bibitem [{\citenamefont {Ivanchenko}(2015)}]{ivanchenko2015quantum}%
  \BibitemOpen
  \bibfield  {author} {\bibinfo {author} {\bibfnamefont {E.~A.}\ \bibnamefont
  {Ivanchenko}},\ }\bibfield  {title} {\enquote {\bibinfo {title} {Quantum otto
  cycle efficiency on coupled qudits},}\ }\href {\doibase
  10.1103/PhysRevE.92.032124} {\bibfield  {journal} {\bibinfo  {journal}
  {Physical Review E}\ }\textbf {\bibinfo {volume} {92}},\ \bibinfo {pages}
  {032124} (\bibinfo {year} {2015})}\BibitemShut {NoStop}%
\bibitem [{\citenamefont {Myers}\ \emph {et~al.}(2021)\citenamefont {Myers},
  \citenamefont {Abah},\ and\ \citenamefont {Deffner}}]{Myers_2021}%
  \BibitemOpen
  \bibfield  {author} {\bibinfo {author} {\bibfnamefont {N.~M.}\ \bibnamefont
  {Myers}}, \bibinfo {author} {\bibfnamefont {O.}~\bibnamefont {Abah}}, \ and\
  \bibinfo {author} {\bibfnamefont {S.}~\bibnamefont {Deffner}},\ }\bibfield
  {title} {\enquote {\bibinfo {title} {Quantum otto engines at relativistic
  energies},}\ }\href {\doibase 10.1088/1367-2630/ac2756} {\bibfield  {journal}
  {\bibinfo  {journal} {New Journal of Physics}\ }\textbf {\bibinfo {volume}
  {23}},\ \bibinfo {pages} {105001} (\bibinfo {year} {2021})}\BibitemShut
  {NoStop}%
\bibitem [{\citenamefont {Wang}\ \emph {et~al.}(2009)\citenamefont {Wang},
  \citenamefont {Liu},\ and\ \citenamefont {He}}]{wang2009performancef}%
  \BibitemOpen
  \bibfield  {author} {\bibinfo {author} {\bibfnamefont {H.}~\bibnamefont
  {Wang}}, \bibinfo {author} {\bibfnamefont {S.}~\bibnamefont {Liu}}, \ and\
  \bibinfo {author} {\bibfnamefont {J.}~\bibnamefont {He}},\ }\bibfield
  {title} {\enquote {\bibinfo {title} {Performance analysis and parametric
  optimum criteria of an irreversible {Bose--Otto} engine},}\ }\href {\doibase
  10.1063/1.3103315} {\bibfield  {journal} {\bibinfo  {journal} {Journal of
  Applied Physics}\ }\textbf {\bibinfo {volume} {105}},\ \bibinfo {pages}
  {083534} (\bibinfo {year} {2009})}\BibitemShut {NoStop}%
\bibitem [{\citenamefont {Myers}\ \emph {et~al.}(2022)\citenamefont {Myers},
  \citenamefont {Pe{\~n}a}, \citenamefont {Negrete}, \citenamefont {Vargas},
  \citenamefont {Chiara},\ and\ \citenamefont {Deffner}}]{myers2022boostingf}%
  \BibitemOpen
  \bibfield  {author} {\bibinfo {author} {\bibfnamefont {N.~M.}\ \bibnamefont
  {Myers}}, \bibinfo {author} {\bibfnamefont {F.~J.}\ \bibnamefont {Pe{\~n}a}},
  \bibinfo {author} {\bibfnamefont {O.}~\bibnamefont {Negrete}}, \bibinfo
  {author} {\bibfnamefont {P.}~\bibnamefont {Vargas}}, \bibinfo {author}
  {\bibfnamefont {G.~De}\ \bibnamefont {Chiara}}, \ and\ \bibinfo {author}
  {\bibfnamefont {S.}~\bibnamefont {Deffner}},\ }\bibfield  {title} {\enquote
  {\bibinfo {title} {Boosting engine performance with {Bose--Einstein}
  condensation},}\ }\href {\doibase 10.1088/1367-2630/ac47cc} {\bibfield
  {journal} {\bibinfo  {journal} {New Journal of Physics}\ }\textbf {\bibinfo
  {volume} {24}},\ \bibinfo {pages} {025001} (\bibinfo {year}
  {2022})}\BibitemShut {NoStop}%
\bibitem [{\citenamefont {Altintas}\ \emph {et~al.}(2015)\citenamefont
  {Altintas}, \citenamefont {Hardal},\ and\ \citenamefont
  {M{\"u}stecapl{\i}o{\u{g}}lu}}]{altintas2015rabi}%
  \BibitemOpen
  \bibfield  {author} {\bibinfo {author} {\bibfnamefont {F.}~\bibnamefont
  {Altintas}}, \bibinfo {author} {\bibfnamefont {A.~{\"U}.~C.}\ \bibnamefont
  {Hardal}}, \ and\ \bibinfo {author} {\bibfnamefont {{\"O}.~E.}\ \bibnamefont
  {M{\"u}stecapl{\i}o{\u{g}}lu}},\ }\bibfield  {title} {\enquote {\bibinfo
  {title} {Rabi model as a quantum coherent heat engine: From quantum biology
  to superconducting circuits},}\ }\href {\doibase 10.1103/PhysRevA.91.023816}
  {\bibfield  {journal} {\bibinfo  {journal} {Physical Review A}\ }\textbf
  {\bibinfo {volume} {91}},\ \bibinfo {pages} {023816} (\bibinfo {year}
  {2015})}\BibitemShut {NoStop}%
\bibitem [{\citenamefont {Song}\ \emph {et~al.}(2016)\citenamefont {Song},
  \citenamefont {Singh}, \citenamefont {Zhang}, \citenamefont {Zhang},\ and\
  \citenamefont {Meystre}}]{song2016one}%
  \BibitemOpen
  \bibfield  {author} {\bibinfo {author} {\bibfnamefont {Q.}~\bibnamefont
  {Song}}, \bibinfo {author} {\bibfnamefont {S.}~\bibnamefont {Singh}},
  \bibinfo {author} {\bibfnamefont {K.}~\bibnamefont {Zhang}}, \bibinfo
  {author} {\bibfnamefont {W.}~\bibnamefont {Zhang}}, \ and\ \bibinfo {author}
  {\bibfnamefont {P.}~\bibnamefont {Meystre}},\ }\bibfield  {title} {\enquote
  {\bibinfo {title} {One qubit and one photon: The simplest polaritonic heat
  engine},}\ }\href {\doibase 10.1103/PhysRevA.94.063852} {\bibfield  {journal}
  {\bibinfo  {journal} {Physical Review A}\ }\textbf {\bibinfo {volume} {94}},\
  \bibinfo {pages} {063852} (\bibinfo {year} {2016})}\BibitemShut {NoStop}%
\bibitem [{\citenamefont {Barrios}\ \emph {et~al.}(2017)\citenamefont
  {Barrios}, \citenamefont {Albarr{\'a}n-Arriagada}, \citenamefont
  {C{\'a}rdenas-L{\'o}pez}, \citenamefont {Romero},\ and\ \citenamefont
  {Retamal}}]{barrios2017role}%
  \BibitemOpen
  \bibfield  {author} {\bibinfo {author} {\bibfnamefont {G.~A.}\ \bibnamefont
  {Barrios}}, \bibinfo {author} {\bibfnamefont {F.}~\bibnamefont
  {Albarr{\'a}n-Arriagada}}, \bibinfo {author} {\bibfnamefont {F.~A.}\
  \bibnamefont {C{\'a}rdenas-L{\'o}pez}}, \bibinfo {author} {\bibfnamefont
  {G.}~\bibnamefont {Romero}}, \ and\ \bibinfo {author} {\bibfnamefont {J.~C.}\
  \bibnamefont {Retamal}},\ }\bibfield  {title} {\enquote {\bibinfo {title}
  {Role of quantum correlations in light-matter quantum heat engines},}\ }\href
  {\doibase 10.1103/PhysRevA.96.052119} {\bibfield  {journal} {\bibinfo
  {journal} {Physical Review A}\ }\textbf {\bibinfo {volume} {96}},\ \bibinfo
  {pages} {052119} (\bibinfo {year} {2017})}\BibitemShut {NoStop}%
\bibitem [{\citenamefont {Mojaveri}\ \emph {et~al.}(2021)\citenamefont
  {Mojaveri}, \citenamefont {Dehghani},\ and\ \citenamefont
  {Ahmadi}}]{mojaveri2021quantumf}%
  \BibitemOpen
  \bibfield  {author} {\bibinfo {author} {\bibfnamefont {B.}~\bibnamefont
  {Mojaveri}}, \bibinfo {author} {\bibfnamefont {A.}~\bibnamefont {Dehghani}},
  \ and\ \bibinfo {author} {\bibfnamefont {Z.}~\bibnamefont {Ahmadi}},\
  }\bibfield  {title} {\enquote {\bibinfo {title} {A quantum correlated heat
  engine based on the parity-deformed {Jaynes--Cummings} model: Achieving the
  classical {Carnot} efficiency by a local classical field},}\ }\href {\doibase
  10.1088/1402-4896/ac1638} {\bibfield  {journal} {\bibinfo  {journal} {Physica
  Scripta}\ }\textbf {\bibinfo {volume} {96}},\ \bibinfo {pages} {115102}
  (\bibinfo {year} {2021})}\BibitemShut {NoStop}%
\bibitem [{\citenamefont {Jaynes}\ and\ \citenamefont
  {Cummings}(1963)}]{jaynes1963comparison}%
  \BibitemOpen
  \bibfield  {author} {\bibinfo {author} {\bibfnamefont {E.~T.}\ \bibnamefont
  {Jaynes}}\ and\ \bibinfo {author} {\bibfnamefont {F.~W.}\ \bibnamefont
  {Cummings}},\ }\bibfield  {title} {\enquote {\bibinfo {title} {Comparison of
  quantum and semiclassical radiation theories with application to the beam
  maser},}\ }\href {\doibase 10.1109/PROC.1963.1664} {\bibfield  {journal}
  {\bibinfo  {journal} {Proceedings of the IEEE}\ }\textbf {\bibinfo {volume}
  {51}},\ \bibinfo {pages} {89--109} (\bibinfo {year} {1963})}\BibitemShut
  {NoStop}%
\bibitem [{\citenamefont {Braak}(2011)}]{braak2011integrability}%
  \BibitemOpen
  \bibfield  {author} {\bibinfo {author} {\bibfnamefont {D.}~\bibnamefont
  {Braak}},\ }\bibfield  {title} {\enquote {\bibinfo {title} {Integrability of
  the rabi model},}\ }\href {\doibase 10.1103/PhysRevLett.107.100401}
  {\bibfield  {journal} {\bibinfo  {journal} {Physical Review Letters}\
  }\textbf {\bibinfo {volume} {107}},\ \bibinfo {pages} {100401} (\bibinfo
  {year} {2011})}\BibitemShut {NoStop}%
\bibitem [{\citenamefont {Chen}\ \emph {et~al.}(2012)\citenamefont {Chen},
  \citenamefont {Wang}, \citenamefont {He}, \citenamefont {Liu},\ and\
  \citenamefont {Wang}}]{chen2012exact}%
  \BibitemOpen
  \bibfield  {author} {\bibinfo {author} {\bibfnamefont {Q.~H.}\ \bibnamefont
  {Chen}}, \bibinfo {author} {\bibfnamefont {C.}~\bibnamefont {Wang}}, \bibinfo
  {author} {\bibfnamefont {S.}~\bibnamefont {He}}, \bibinfo {author}
  {\bibfnamefont {T.}~\bibnamefont {Liu}}, \ and\ \bibinfo {author}
  {\bibfnamefont {K.~L.}\ \bibnamefont {Wang}},\ }\bibfield  {title} {\enquote
  {\bibinfo {title} {Exact solvability of the quantum rabi model using
  bogoliubov operators},}\ }\href {\doibase 10.1103/PhysRevA.86.023822}
  {\bibfield  {journal} {\bibinfo  {journal} {Physical Review A}\ }\textbf
  {\bibinfo {volume} {86}},\ \bibinfo {pages} {023822} (\bibinfo {year}
  {2012})}\BibitemShut {NoStop}%
\bibitem [{\citenamefont {Braak}\ \emph {et~al.}(2016)\citenamefont {Braak},
  \citenamefont {Chen}, \citenamefont {Batchelor},\ and\ \citenamefont
  {Solano}}]{braak2016semi}%
  \BibitemOpen
  \bibfield  {author} {\bibinfo {author} {\bibfnamefont {D.}~\bibnamefont
  {Braak}}, \bibinfo {author} {\bibfnamefont {Q.~H.}\ \bibnamefont {Chen}},
  \bibinfo {author} {\bibfnamefont {M.~T.}\ \bibnamefont {Batchelor}}, \ and\
  \bibinfo {author} {\bibfnamefont {E.}~\bibnamefont {Solano}},\ }\bibfield
  {title} {\enquote {\bibinfo {title} {Semi-classical and quantum rabi models:
  in celebration of 80 years},}\ }\href {\doibase
  10.1088/1751-8113/49/30/300301} {\bibfield  {journal} {\bibinfo  {journal}
  {Journal of Physics A: Mathematical and Theoretical}\ }\textbf {\bibinfo
  {volume} {49}},\ \bibinfo {pages} {300301} (\bibinfo {year}
  {2016})}\BibitemShut {NoStop}%
\bibitem [{\citenamefont {Rabi}(1936)}]{rabi1936process}%
  \BibitemOpen
  \bibfield  {author} {\bibinfo {author} {\bibfnamefont {I.~I.}\ \bibnamefont
  {Rabi}},\ }\bibfield  {title} {\enquote {\bibinfo {title} {On the process of
  space quantization},}\ }\href {\doibase 10.1103/PhysRev.49.324} {\bibfield
  {journal} {\bibinfo  {journal} {Physical Review}\ }\textbf {\bibinfo {volume}
  {49}},\ \bibinfo {pages} {324} (\bibinfo {year} {1936})}\BibitemShut
  {NoStop}%
\bibitem [{\citenamefont {Rabi}(1937)}]{rabi1937space}%
  \BibitemOpen
  \bibfield  {author} {\bibinfo {author} {\bibfnamefont {I.~I.}\ \bibnamefont
  {Rabi}},\ }\bibfield  {title} {\enquote {\bibinfo {title} {Space quantization
  in a gyrating magnetic field},}\ }\href {\doibase 10.1103/PhysRev.51.652}
  {\bibfield  {journal} {\bibinfo  {journal} {Physical Review}\ }\textbf
  {\bibinfo {volume} {51}},\ \bibinfo {pages} {652} (\bibinfo {year}
  {1937})}\BibitemShut {NoStop}%
\bibitem [{\citenamefont {Xu}\ \emph {et~al.}(2024{\natexlab{a}})\citenamefont
  {Xu}, \citenamefont {Jin}, \citenamefont {de~Almeida},\ and\ \citenamefont
  {de~Moraes~Neto}}]{xu2024exploring}%
  \BibitemOpen
  \bibfield  {author} {\bibinfo {author} {\bibfnamefont {H.~G.}\ \bibnamefont
  {Xu}}, \bibinfo {author} {\bibfnamefont {J.}~\bibnamefont {Jin}}, \bibinfo
  {author} {\bibfnamefont {N.~G.}\ \bibnamefont {de~Almeida}}, \ and\ \bibinfo
  {author} {\bibfnamefont {G.~D.}\ \bibnamefont {de~Moraes~Neto}},\ }\bibfield
  {title} {\enquote {\bibinfo {title} {Exploring the role of criticality in the
  quantum otto cycle fueled by the anisotropic quantum rabi-stark model},}\
  }\href {\doibase 10.1103/PhysRevB.110.134318} {\bibfield  {journal} {\bibinfo
   {journal} {Physical Review B}\ }\textbf {\bibinfo {volume} {110}},\ \bibinfo
  {pages} {134318} (\bibinfo {year} {2024}{\natexlab{a}})}\BibitemShut
  {NoStop}%
\bibitem [{\citenamefont {Xu}\ \emph {et~al.}(2024{\natexlab{b}})\citenamefont
  {Xu}, \citenamefont {Jin}, \citenamefont {Neto},\ and\ \citenamefont
  {de~Almeida}}]{xu2024universal}%
  \BibitemOpen
  \bibfield  {author} {\bibinfo {author} {\bibfnamefont {H.~G.}\ \bibnamefont
  {Xu}}, \bibinfo {author} {\bibfnamefont {J.}~\bibnamefont {Jin}}, \bibinfo
  {author} {\bibfnamefont {G.~D.~M.}\ \bibnamefont {Neto}}, \ and\ \bibinfo
  {author} {\bibfnamefont {N.~G.}\ \bibnamefont {de~Almeida}},\ }\bibfield
  {title} {\enquote {\bibinfo {title} {Universal quantum otto heat machine
  based on the dicke model},}\ }\href {\doibase 10.1103/PhysRevE.109.014122}
  {\bibfield  {journal} {\bibinfo  {journal} {Physical Review E}\ }\textbf
  {\bibinfo {volume} {109}},\ \bibinfo {pages} {014122} (\bibinfo {year}
  {2024}{\natexlab{b}})}\BibitemShut {NoStop}%
\bibitem [{\citenamefont {Quan}\ \emph {et~al.}(2006)\citenamefont {Quan},
  \citenamefont {Wang}, \citenamefont {Liu}, \citenamefont {Sun},\ and\
  \citenamefont {Nori}}]{quan2006maxwell}%
  \BibitemOpen
  \bibfield  {author} {\bibinfo {author} {\bibfnamefont {H.~T.}\ \bibnamefont
  {Quan}}, \bibinfo {author} {\bibfnamefont {Y.~D.}\ \bibnamefont {Wang}},
  \bibinfo {author} {\bibfnamefont {Y.~X.}\ \bibnamefont {Liu}}, \bibinfo
  {author} {\bibfnamefont {C.~P.}\ \bibnamefont {Sun}}, \ and\ \bibinfo
  {author} {\bibfnamefont {F.}~\bibnamefont {Nori}},\ }\bibfield  {title}
  {\enquote {\bibinfo {title} {Maxwell's demon assisted thermodynamic cycle in
  superconducting quantum circuits},}\ }\href {\doibase
  10.1103/PhysRevLett.97.180402} {\bibfield  {journal} {\bibinfo  {journal}
  {Physical review letters}\ }\textbf {\bibinfo {volume} {97}},\ \bibinfo
  {pages} {180402} (\bibinfo {year} {2006})}\BibitemShut {NoStop}%
\bibitem [{\citenamefont {Niskanen}\ \emph {et~al.}(2007)\citenamefont
  {Niskanen}, \citenamefont {Nakamura},\ and\ \citenamefont
  {Pekola}}]{niskanen2007information}%
  \BibitemOpen
  \bibfield  {author} {\bibinfo {author} {\bibfnamefont {A.~O.}\ \bibnamefont
  {Niskanen}}, \bibinfo {author} {\bibfnamefont {Y.}~\bibnamefont {Nakamura}},
  \ and\ \bibinfo {author} {\bibfnamefont {J.~P.}\ \bibnamefont {Pekola}},\
  }\bibfield  {title} {\enquote {\bibinfo {title} {Information entropic
  superconducting microcooler},}\ }\href {\doibase 10.1103/PhysRevB.76.174523}
  {\bibfield  {journal} {\bibinfo  {journal} {Physical Review B}\ }\textbf
  {\bibinfo {volume} {76}},\ \bibinfo {pages} {174523} (\bibinfo {year}
  {2007})}\BibitemShut {NoStop}%
\bibitem [{\citenamefont {Pekola}\ \emph {et~al.}(2010)\citenamefont {Pekola},
  \citenamefont {Brosco}, \citenamefont {M{\"o}tt{\"o}nen}, \citenamefont
  {Solinas},\ and\ \citenamefont {Shnirman}}]{pekola2010decoherence}%
  \BibitemOpen
  \bibfield  {author} {\bibinfo {author} {\bibfnamefont {J.~P.}\ \bibnamefont
  {Pekola}}, \bibinfo {author} {\bibfnamefont {V.}~\bibnamefont {Brosco}},
  \bibinfo {author} {\bibfnamefont {M.}~\bibnamefont {M{\"o}tt{\"o}nen}},
  \bibinfo {author} {\bibfnamefont {P.}~\bibnamefont {Solinas}}, \ and\
  \bibinfo {author} {\bibfnamefont {A.}~\bibnamefont {Shnirman}},\ }\bibfield
  {title} {\enquote {\bibinfo {title} {Decoherence in adiabatic quantum
  evolution: Application to cooper pair pumping},}\ }\href {\doibase
  10.1103/PhysRevLett.105.030401} {\bibfield  {journal} {\bibinfo  {journal}
  {Physical review letters}\ }\textbf {\bibinfo {volume} {105}},\ \bibinfo
  {pages} {030401} (\bibinfo {year} {2010})}\BibitemShut {NoStop}%
\bibitem [{\citenamefont {Koski}\ \emph {et~al.}(2015)\citenamefont {Koski},
  \citenamefont {Kutvonen}, \citenamefont {Khaymovich}, \citenamefont
  {Ala-Nissila},\ and\ \citenamefont {Pekola}}]{koski2015chip}%
  \BibitemOpen
  \bibfield  {author} {\bibinfo {author} {\bibfnamefont {J.~V.}\ \bibnamefont
  {Koski}}, \bibinfo {author} {\bibfnamefont {A.}~\bibnamefont {Kutvonen}},
  \bibinfo {author} {\bibfnamefont {I.~M.}\ \bibnamefont {Khaymovich}},
  \bibinfo {author} {\bibfnamefont {T.}~\bibnamefont {Ala-Nissila}}, \ and\
  \bibinfo {author} {\bibfnamefont {J.~P.}\ \bibnamefont {Pekola}},\ }\bibfield
   {title} {\enquote {\bibinfo {title} {On-chip maxwell's demon as an
  information-powered refrigerator},}\ }\href {\doibase
  10.1103/PhysRevLett.115.260602} {\bibfield  {journal} {\bibinfo  {journal}
  {Physical review letters}\ }\textbf {\bibinfo {volume} {115}},\ \bibinfo
  {pages} {260602} (\bibinfo {year} {2015})}\BibitemShut {NoStop}%
\bibitem [{\citenamefont {Pekola}\ \emph {et~al.}(2016)\citenamefont {Pekola},
  \citenamefont {Golubev},\ and\ \citenamefont {Averin}}]{pekola2016maxwell}%
  \BibitemOpen
  \bibfield  {author} {\bibinfo {author} {\bibfnamefont {J.~P.}\ \bibnamefont
  {Pekola}}, \bibinfo {author} {\bibfnamefont {D.~S.}\ \bibnamefont {Golubev}},
  \ and\ \bibinfo {author} {\bibfnamefont {D.~V.}\ \bibnamefont {Averin}},\
  }\bibfield  {title} {\enquote {\bibinfo {title} {Maxwell's demon based on a
  single qubit},}\ }\href {\doibase 10.1103/PhysRevB.93.024501} {\bibfield
  {journal} {\bibinfo  {journal} {Physical Review B}\ }\textbf {\bibinfo
  {volume} {93}},\ \bibinfo {pages} {024501} (\bibinfo {year}
  {2016})}\BibitemShut {NoStop}%
\bibitem [{\citenamefont {Ro{\ss}nagel}\ \emph
  {et~al.}(2014{\natexlab{b}})\citenamefont {Ro{\ss}nagel}, \citenamefont
  {Abah}, \citenamefont {Schmidt-Kaler}, \citenamefont {Singer},\ and\
  \citenamefont {Lutz}}]{rossnagel2014nanoscale}%
  \BibitemOpen
  \bibfield  {author} {\bibinfo {author} {\bibfnamefont {J.}~\bibnamefont
  {Ro{\ss}nagel}}, \bibinfo {author} {\bibfnamefont {O.}~\bibnamefont {Abah}},
  \bibinfo {author} {\bibfnamefont {F.}~\bibnamefont {Schmidt-Kaler}}, \bibinfo
  {author} {\bibfnamefont {K.}~\bibnamefont {Singer}}, \ and\ \bibinfo {author}
  {\bibfnamefont {E.}~\bibnamefont {Lutz}},\ }\bibfield  {title} {\enquote
  {\bibinfo {title} {Nanoscale heat engine beyond the carnot limit},}\ }\href
  {\doibase 10.1103/PhysRevLett.112.030602} {\bibfield  {journal} {\bibinfo
  {journal} {Physical review letters}\ }\textbf {\bibinfo {volume} {112}},\
  \bibinfo {pages} {030602} (\bibinfo {year} {2014}{\natexlab{b}})}\BibitemShut
  {NoStop}%
\bibitem [{\citenamefont {Ro{\ss}nagel}\ \emph {et~al.}(2016)\citenamefont
  {Ro{\ss}nagel}, \citenamefont {Dawkins}, \citenamefont {Tolazzi},
  \citenamefont {Abah}, \citenamefont {Lutz}, \citenamefont {Schmidt-Kaler},\
  and\ \citenamefont {Singer}}]{rossnagel2016single}%
  \BibitemOpen
  \bibfield  {author} {\bibinfo {author} {\bibfnamefont {J.}~\bibnamefont
  {Ro{\ss}nagel}}, \bibinfo {author} {\bibfnamefont {S.~T.}\ \bibnamefont
  {Dawkins}}, \bibinfo {author} {\bibfnamefont {K.~N.}\ \bibnamefont
  {Tolazzi}}, \bibinfo {author} {\bibfnamefont {O.}~\bibnamefont {Abah}},
  \bibinfo {author} {\bibfnamefont {E.}~\bibnamefont {Lutz}}, \bibinfo {author}
  {\bibfnamefont {F.}~\bibnamefont {Schmidt-Kaler}}, \ and\ \bibinfo {author}
  {\bibfnamefont {K.}~\bibnamefont {Singer}},\ }\bibfield  {title} {\enquote
  {\bibinfo {title} {A single-atom heat engine},}\ }\href {\doibase
  10.1126/science.aad6320} {\bibfield  {journal} {\bibinfo  {journal}
  {Science}\ }\textbf {\bibinfo {volume} {352}},\ \bibinfo {pages} {325--329}
  (\bibinfo {year} {2016})}\BibitemShut {NoStop}%
\bibitem [{\citenamefont {Maslennikov}\ \emph {et~al.}(2019)\citenamefont
  {Maslennikov}, \citenamefont {Ding}, \citenamefont {Habl{\"u}tzel},
  \citenamefont {Gan}, \citenamefont {Roulet}, \citenamefont {Nimmrichter},
  \citenamefont {Dai}, \citenamefont {Scarani},\ and\ \citenamefont
  {Matsukevich}}]{maslennikov2019quantumf}%
  \BibitemOpen
  \bibfield  {author} {\bibinfo {author} {\bibfnamefont {G.}~\bibnamefont
  {Maslennikov}}, \bibinfo {author} {\bibfnamefont {S.}~\bibnamefont {Ding}},
  \bibinfo {author} {\bibfnamefont {R.}~\bibnamefont {Habl{\"u}tzel}}, \bibinfo
  {author} {\bibfnamefont {J.}~\bibnamefont {Gan}}, \bibinfo {author}
  {\bibfnamefont {A.}~\bibnamefont {Roulet}}, \bibinfo {author} {\bibfnamefont
  {S.}~\bibnamefont {Nimmrichter}}, \bibinfo {author} {\bibfnamefont
  {J.}~\bibnamefont {Dai}}, \bibinfo {author} {\bibfnamefont {V.}~\bibnamefont
  {Scarani}}, \ and\ \bibinfo {author} {\bibfnamefont {D.}~\bibnamefont
  {Matsukevich}},\ }\bibfield  {title} {\enquote {\bibinfo {title} {Quantum
  absorption refrigerator with trapped ions},}\ }\href {\doibase
  10.1038/s41467-018-08090-0} {\bibfield  {journal} {\bibinfo  {journal}
  {Nature Communications}\ }\textbf {\bibinfo {volume} {10}},\ \bibinfo {pages}
  {202} (\bibinfo {year} {2019})}\BibitemShut {NoStop}%
\bibitem [{\citenamefont {Zhang}\ \emph
  {et~al.}(2014{\natexlab{b}})\citenamefont {Zhang}, \citenamefont {Bariani},\
  and\ \citenamefont {Meystre}}]{zhang2014quantum}%
  \BibitemOpen
  \bibfield  {author} {\bibinfo {author} {\bibfnamefont {K.}~\bibnamefont
  {Zhang}}, \bibinfo {author} {\bibfnamefont {F.}~\bibnamefont {Bariani}}, \
  and\ \bibinfo {author} {\bibfnamefont {P.}~\bibnamefont {Meystre}},\
  }\bibfield  {title} {\enquote {\bibinfo {title} {Quantum optomechanical heat
  engine},}\ }\href {\doibase 10.1103/PhysRevLett.112.150602} {\bibfield
  {journal} {\bibinfo  {journal} {Physical review letters}\ }\textbf {\bibinfo
  {volume} {112}},\ \bibinfo {pages} {150602} (\bibinfo {year}
  {2014}{\natexlab{b}})}\BibitemShut {NoStop}%
\bibitem [{\citenamefont {Zhang}\ \emph
  {et~al.}(2014{\natexlab{c}})\citenamefont {Zhang}, \citenamefont {Bariani},\
  and\ \citenamefont {Meystre}}]{2014Theory}%
  \BibitemOpen
  \bibfield  {author} {\bibinfo {author} {\bibfnamefont {K.}~\bibnamefont
  {Zhang}}, \bibinfo {author} {\bibfnamefont {F.}~\bibnamefont {Bariani}}, \
  and\ \bibinfo {author} {\bibfnamefont {P.}~\bibnamefont {Meystre}},\
  }\bibfield  {title} {\enquote {\bibinfo {title} {Theory of an optomechanical
  quantum heat engine},}\ }\href {\doibase 10.1103/PhysRevA.90.023819}
  {\bibfield  {journal} {\bibinfo  {journal} {Physical Review A}\ }\textbf
  {\bibinfo {volume} {90}},\ \bibinfo {pages} {023819} (\bibinfo {year}
  {2014}{\natexlab{c}})}\BibitemShut {NoStop}%
\bibitem [{\citenamefont {Fialko}\ and\ \citenamefont
  {Hallwood}(2012)}]{2012Isolated}%
  \BibitemOpen
  \bibfield  {author} {\bibinfo {author} {\bibfnamefont {O.}~\bibnamefont
  {Fialko}}\ and\ \bibinfo {author} {\bibfnamefont {D.}~\bibnamefont
  {Hallwood}},\ }\bibfield  {title} {\enquote {\bibinfo {title} {Isolated
  quantum heat engine},}\ }\href {\doibase 10.1103/PhysRevLett.108.085303}
  {\bibfield  {journal} {\bibinfo  {journal} {Physical Review Letters}\
  }\textbf {\bibinfo {volume} {108}},\ \bibinfo {pages} {085303} (\bibinfo
  {year} {2012})}\BibitemShut {NoStop}%
\bibitem [{\citenamefont {Brantut}\ \emph {et~al.}(2013)\citenamefont
  {Brantut}, \citenamefont {Grenier}, \citenamefont {Meineke}, \citenamefont
  {Stadler}, \citenamefont {Krinner}, \citenamefont {Kollath}, \citenamefont
  {Esslinger},\ and\ \citenamefont {Georges}}]{brantut2013thermoelectric}%
  \BibitemOpen
  \bibfield  {author} {\bibinfo {author} {\bibfnamefont {J.~P.}\ \bibnamefont
  {Brantut}}, \bibinfo {author} {\bibfnamefont {C.}~\bibnamefont {Grenier}},
  \bibinfo {author} {\bibfnamefont {J.}~\bibnamefont {Meineke}}, \bibinfo
  {author} {\bibfnamefont {D.}~\bibnamefont {Stadler}}, \bibinfo {author}
  {\bibfnamefont {S.}~\bibnamefont {Krinner}}, \bibinfo {author} {\bibfnamefont
  {C.}~\bibnamefont {Kollath}}, \bibinfo {author} {\bibfnamefont
  {T.}~\bibnamefont {Esslinger}}, \ and\ \bibinfo {author} {\bibfnamefont
  {A.}~\bibnamefont {Georges}},\ }\bibfield  {title} {\enquote {\bibinfo
  {title} {A thermoelectric heat engine with ultracold atoms},}\ }\href
  {\doibase 10.1126/science.1242308} {\bibfield  {journal} {\bibinfo  {journal}
  {Science}\ }\textbf {\bibinfo {volume} {342}},\ \bibinfo {pages} {713--715}
  (\bibinfo {year} {2013})}\BibitemShut {NoStop}%
\bibitem [{\citenamefont {Batalh{\~a}o}\ \emph {et~al.}(2014)\citenamefont
  {Batalh{\~a}o}, \citenamefont {Souza}, \citenamefont {Mazzola}, \citenamefont
  {Auccaise}, \citenamefont {Sarthour}, \citenamefont {Oliveira}, \citenamefont
  {Goold}, \citenamefont {Chiara}, \citenamefont {Paternostro},\ and\
  \citenamefont {Serra}}]{batalhao2014experimental}%
  \BibitemOpen
  \bibfield  {author} {\bibinfo {author} {\bibfnamefont {T.~B.}\ \bibnamefont
  {Batalh{\~a}o}}, \bibinfo {author} {\bibfnamefont {A.~M.}\ \bibnamefont
  {Souza}}, \bibinfo {author} {\bibfnamefont {L.}~\bibnamefont {Mazzola}},
  \bibinfo {author} {\bibfnamefont {R.}~\bibnamefont {Auccaise}}, \bibinfo
  {author} {\bibfnamefont {R.~S.}\ \bibnamefont {Sarthour}}, \bibinfo {author}
  {\bibfnamefont {I.~S.}\ \bibnamefont {Oliveira}}, \bibinfo {author}
  {\bibfnamefont {J.}~\bibnamefont {Goold}}, \bibinfo {author} {\bibfnamefont
  {G.~De}\ \bibnamefont {Chiara}}, \bibinfo {author} {\bibfnamefont
  {M.}~\bibnamefont {Paternostro}}, \ and\ \bibinfo {author} {\bibfnamefont
  {R.~M.}\ \bibnamefont {Serra}},\ }\bibfield  {title} {\enquote {\bibinfo
  {title} {Experimental reconstruction of work distribution and study of
  fluctuation relations in a closed quantum system},}\ }\href {\doibase
  10.1103/PhysRevLett.113.140601} {\bibfield  {journal} {\bibinfo  {journal}
  {Physical review letters}\ }\textbf {\bibinfo {volume} {113}},\ \bibinfo
  {pages} {140601} (\bibinfo {year} {2014})}\BibitemShut {NoStop}%
\bibitem [{\citenamefont {Micadei}\ \emph {et~al.}(2019)\citenamefont
  {Micadei}, \citenamefont {Peterson}, \citenamefont {Souza}, \citenamefont
  {Sarthour}, \citenamefont {Oliveira}, \citenamefont {Landi}, \citenamefont
  {Batalh{\~a}o}, \citenamefont {Serra},\ and\ \citenamefont
  {Lutz}}]{micadei2019reversing}%
  \BibitemOpen
  \bibfield  {author} {\bibinfo {author} {\bibfnamefont {K.}~\bibnamefont
  {Micadei}}, \bibinfo {author} {\bibfnamefont {J.~P.~S.}\ \bibnamefont
  {Peterson}}, \bibinfo {author} {\bibfnamefont {A.~M.}\ \bibnamefont {Souza}},
  \bibinfo {author} {\bibfnamefont {R.~S.}\ \bibnamefont {Sarthour}}, \bibinfo
  {author} {\bibfnamefont {I.~S.}\ \bibnamefont {Oliveira}}, \bibinfo {author}
  {\bibfnamefont {G.~T.}\ \bibnamefont {Landi}}, \bibinfo {author}
  {\bibfnamefont {T.~B.}\ \bibnamefont {Batalh{\~a}o}}, \bibinfo {author}
  {\bibfnamefont {R.~M.}\ \bibnamefont {Serra}}, \ and\ \bibinfo {author}
  {\bibfnamefont {E.}~\bibnamefont {Lutz}},\ }\bibfield  {title} {\enquote
  {\bibinfo {title} {Reversing the direction of heat flow using quantum
  correlations},}\ }\href {\doibase 10.1038/s41467-019-10333-7} {\bibfield
  {journal} {\bibinfo  {journal} {Nature communications}\ }\textbf {\bibinfo
  {volume} {10}},\ \bibinfo {pages} {2456} (\bibinfo {year}
  {2019})}\BibitemShut {NoStop}%
\bibitem [{\citenamefont {de~Assis}\ \emph {et~al.}(2019)\citenamefont
  {de~Assis}, \citenamefont {de~Mendon{\c{c}}a}, \citenamefont {Villas-Boas},
  \citenamefont {de~Souza}, \citenamefont {Sarthour}, \citenamefont
  {Oliveira},\ and\ \citenamefont {de~Almeida}}]{de2019efficiency}%
  \BibitemOpen
  \bibfield  {author} {\bibinfo {author} {\bibfnamefont {R.~J.}\ \bibnamefont
  {de~Assis}}, \bibinfo {author} {\bibfnamefont {T.~M.}\ \bibnamefont
  {de~Mendon{\c{c}}a}}, \bibinfo {author} {\bibfnamefont {C.~J.}\ \bibnamefont
  {Villas-Boas}}, \bibinfo {author} {\bibfnamefont {A.~M.}\ \bibnamefont
  {de~Souza}}, \bibinfo {author} {\bibfnamefont {R.~S.}\ \bibnamefont
  {Sarthour}}, \bibinfo {author} {\bibfnamefont {I.~S.}\ \bibnamefont
  {Oliveira}}, \ and\ \bibinfo {author} {\bibfnamefont {N.~G.}\ \bibnamefont
  {de~Almeida}},\ }\bibfield  {title} {\enquote {\bibinfo {title} {Efficiency
  of a quantum otto heat engine operating under a reservoir at effective
  negative temperatures},}\ }\href {\doibase 10.1103/PhysRevLett.122.240602}
  {\bibfield  {journal} {\bibinfo  {journal} {Physical review letters}\
  }\textbf {\bibinfo {volume} {122}},\ \bibinfo {pages} {240602} (\bibinfo
  {year} {2019})}\BibitemShut {NoStop}%
\bibitem [{\citenamefont {Bouton}\ \emph {et~al.}(2021)\citenamefont {Bouton},
  \citenamefont {Nettersheim}, \citenamefont {Adam}, \citenamefont {Lausch},
  \citenamefont {Mayer}, \citenamefont {Schmidt},\ and\ \citenamefont
  {Widera}}]{Bouton2021}%
  \BibitemOpen
  \bibfield  {author} {\bibinfo {author} {\bibfnamefont {Q.}~\bibnamefont
  {Bouton}}, \bibinfo {author} {\bibfnamefont {J.}~\bibnamefont {Nettersheim}},
  \bibinfo {author} {\bibfnamefont {D.}~\bibnamefont {Adam}}, \bibinfo {author}
  {\bibfnamefont {T.}~\bibnamefont {Lausch}}, \bibinfo {author} {\bibfnamefont
  {F.}~\bibnamefont {Mayer}}, \bibinfo {author} {\bibfnamefont
  {F.}~\bibnamefont {Schmidt}}, \ and\ \bibinfo {author} {\bibfnamefont
  {A.}~\bibnamefont {Widera}},\ }\bibfield  {title} {\enquote {\bibinfo {title}
  {A quantum heat engine driven by atomic collisions},}\ }\href {\doibase
  10.1038/s41467-021-22222-z} {\bibfield  {journal} {\bibinfo  {journal}
  {Nature Communications}\ }\textbf {\bibinfo {volume} {12}},\ \bibinfo {pages}
  {2063} (\bibinfo {year} {2021})}\BibitemShut {NoStop}%
\bibitem [{\citenamefont {Alsulami}\ and\ \citenamefont
  {Abd-Rabbou}(2024)}]{Alsulami2024tow}%
  \BibitemOpen
  \bibfield  {author} {\bibinfo {author} {\bibfnamefont {M.~D.}\ \bibnamefont
  {Alsulami}}\ and\ \bibinfo {author} {\bibfnamefont {M.~Y.}\ \bibnamefont
  {Abd-Rabbou}},\ }\bibfield  {title} {\enquote {\bibinfo {title} {{Quantum
  Heat Engines with Spin-Chain-Star Systems}},}\ }\href {\doibase
  10.1002/andp.202400122} {\bibfield  {journal} {\bibinfo  {journal} {Annalen
  Phys.}\ }\textbf {\bibinfo {volume} {536}},\ \bibinfo {pages} {2400122}
  (\bibinfo {year} {2024})}\BibitemShut {NoStop}%
\bibitem [{\citenamefont {Hardal}\ and\ \citenamefont
  {M{\"u}stecapl{\i}o{\u{g}}lu}(2015)}]{hardal2015superradiant}%
  \BibitemOpen
  \bibfield  {author} {\bibinfo {author} {\bibfnamefont {A.~{\"U}.~C.}\
  \bibnamefont {Hardal}}\ and\ \bibinfo {author} {\bibfnamefont {{\"O}.~E.}\
  \bibnamefont {M{\"u}stecapl{\i}o{\u{g}}lu}},\ }\bibfield  {title} {\enquote
  {\bibinfo {title} {Superradiant quantum heat engine},}\ }\href {\doibase
  10.1038/srep12953} {\bibfield  {journal} {\bibinfo  {journal} {Scientific
  Reports}\ }\textbf {\bibinfo {volume} {5}},\ \bibinfo {pages} {12953}
  (\bibinfo {year} {2015})}\BibitemShut {NoStop}%
\bibitem [{\citenamefont {Jaramillo}\ \emph {et~al.}(2016)\citenamefont
  {Jaramillo}, \citenamefont {Beau},\ and\ \citenamefont {del
  Campo}}]{Jaramillo2016}%
  \BibitemOpen
  \bibfield  {author} {\bibinfo {author} {\bibfnamefont {J.}~\bibnamefont
  {Jaramillo}}, \bibinfo {author} {\bibfnamefont {M.}~\bibnamefont {Beau}}, \
  and\ \bibinfo {author} {\bibfnamefont {A.}~\bibnamefont {del Campo}},\
  }\bibfield  {title} {\enquote {\bibinfo {title} {Quantum supremacy of
  many-particle thermal machines},}\ }\href {\doibase
  10.1088/1367-2630/18/7/075019} {\bibfield  {journal} {\bibinfo  {journal}
  {New Journal of Physics}\ }\textbf {\bibinfo {volume} {18}},\ \bibinfo
  {pages} {075019} (\bibinfo {year} {2016})}\BibitemShut {NoStop}%
\bibitem [{\citenamefont {Dicke}(1954)}]{Dicke1954}%
  \BibitemOpen
  \bibfield  {author} {\bibinfo {author} {\bibfnamefont {R.~H.}\ \bibnamefont
  {Dicke}},\ }\bibfield  {title} {\enquote {\bibinfo {title} {Coherence in
  spontaneous radiation processes},}\ }\href {\doibase 10.1103/PhysRev.93.99}
  {\bibfield  {journal} {\bibinfo  {journal} {Physical Review}\ }\textbf
  {\bibinfo {volume} {93}},\ \bibinfo {pages} {99} (\bibinfo {year}
  {1954})}\BibitemShut {NoStop}%
\bibitem [{\citenamefont {Bhaseen}\ \emph {et~al.}(2012)\citenamefont
  {Bhaseen}, \citenamefont {Mayoh}, \citenamefont {Simons},\ and\ \citenamefont
  {Keeling}}]{Bhaseen2012}%
  \BibitemOpen
  \bibfield  {author} {\bibinfo {author} {\bibfnamefont {M.~J.}\ \bibnamefont
  {Bhaseen}}, \bibinfo {author} {\bibfnamefont {J.}~\bibnamefont {Mayoh}},
  \bibinfo {author} {\bibfnamefont {B.~D.}\ \bibnamefont {Simons}}, \ and\
  \bibinfo {author} {\bibfnamefont {J.}~\bibnamefont {Keeling}},\ }\bibfield
  {title} {\enquote {\bibinfo {title} {Dynamics of nonequilibrium {Dicke}
  models},}\ }\href {\doibase 10.1103/PhysRevA.85.013817} {\bibfield  {journal}
  {\bibinfo  {journal} {Physical Review A}\ }\textbf {\bibinfo {volume} {85}},\
  \bibinfo {pages} {013817} (\bibinfo {year} {2012})}\BibitemShut {NoStop}%
\bibitem [{\citenamefont {Greiner}\ \emph {et~al.}(2002)\citenamefont
  {Greiner}, \citenamefont {Mandel}, \citenamefont {Esslinger}, \citenamefont
  {H{\"a}nsch},\ and\ \citenamefont {Bloch}}]{Greiner2002}%
  \BibitemOpen
  \bibfield  {author} {\bibinfo {author} {\bibfnamefont {M.}~\bibnamefont
  {Greiner}}, \bibinfo {author} {\bibfnamefont {O.}~\bibnamefont {Mandel}},
  \bibinfo {author} {\bibfnamefont {T.}~\bibnamefont {Esslinger}}, \bibinfo
  {author} {\bibfnamefont {T.~W.}\ \bibnamefont {H{\"a}nsch}}, \ and\ \bibinfo
  {author} {\bibfnamefont {I.}~\bibnamefont {Bloch}},\ }\bibfield  {title}
  {\enquote {\bibinfo {title} {Quantum phase transition from a superfluid to a
  {Mott} insulator in a gas of ultracold atoms},}\ }\href {\doibase
  10.1038/415039a} {\bibfield  {journal} {\bibinfo  {journal} {Nature}\
  }\textbf {\bibinfo {volume} {415}},\ \bibinfo {pages} {39--44} (\bibinfo
  {year} {2002})}\BibitemShut {NoStop}%
\bibitem [{\citenamefont {Hwang}\ \emph {et~al.}(2015)\citenamefont {Hwang},
  \citenamefont {Puebla},\ and\ \citenamefont {Plenio}}]{PhysRevLett115180404}%
  \BibitemOpen
  \bibfield  {author} {\bibinfo {author} {\bibfnamefont {M.~J.}\ \bibnamefont
  {Hwang}}, \bibinfo {author} {\bibfnamefont {R.}~\bibnamefont {Puebla}}, \
  and\ \bibinfo {author} {\bibfnamefont {M.~B.}\ \bibnamefont {Plenio}},\
  }\bibfield  {title} {\enquote {\bibinfo {title} {Quantum phase transition and
  universal dynamics in the rabi model},}\ }\href {\doibase
  10.1103/PhysRevLett.115.180404} {\bibfield  {journal} {\bibinfo  {journal}
  {Phys. Rev. Lett.}\ }\textbf {\bibinfo {volume} {115}},\ \bibinfo {pages}
  {180404} (\bibinfo {year} {2015})}\BibitemShut {NoStop}%
\bibitem [{\citenamefont {Nataf}\ and\ \citenamefont
  {Ciuti}(2010)}]{Nataf2010}%
  \BibitemOpen
  \bibfield  {author} {\bibinfo {author} {\bibfnamefont {P.}~\bibnamefont
  {Nataf}}\ and\ \bibinfo {author} {\bibfnamefont {C.}~\bibnamefont {Ciuti}},\
  }\bibfield  {title} {\enquote {\bibinfo {title} {No-go theorem for
  superradiant quantum phase transitions in cavity {QED} and counter-example in
  circuit {QED}},}\ }\href {\doibase 10.1038/ncomms1069} {\bibfield  {journal}
  {\bibinfo  {journal} {Nature Communications}\ }\textbf {\bibinfo {volume}
  {1}},\ \bibinfo {pages} {72} (\bibinfo {year} {2010})}\BibitemShut {NoStop}%
\bibitem [{\citenamefont {Greentree}\ \emph {et~al.}(2006)\citenamefont
  {Greentree}, \citenamefont {Tahan}, \citenamefont {Cole},\ and\ \citenamefont
  {Hollenberg}}]{greentree2006quantum}%
  \BibitemOpen
  \bibfield  {author} {\bibinfo {author} {\bibfnamefont {A.~D.}\ \bibnamefont
  {Greentree}}, \bibinfo {author} {\bibfnamefont {C.}~\bibnamefont {Tahan}},
  \bibinfo {author} {\bibfnamefont {J.~H.}\ \bibnamefont {Cole}}, \ and\
  \bibinfo {author} {\bibfnamefont {L.~C.~L.}\ \bibnamefont {Hollenberg}},\
  }\bibfield  {title} {\enquote {\bibinfo {title} {Quantum phase transitions of
  light},}\ }\href {\doibase 10.1038/nphys466} {\bibfield  {journal} {\bibinfo
  {journal} {Nature Physics}\ }\textbf {\bibinfo {volume} {2}},\ \bibinfo
  {pages} {856--861} (\bibinfo {year} {2006})}\BibitemShut {NoStop}%
\bibitem [{\citenamefont {Hartmann}\ \emph {et~al.}(2006)\citenamefont
  {Hartmann}, \citenamefont {Brandao},\ and\ \citenamefont
  {Plenio}}]{hartmann2006strongly}%
  \BibitemOpen
  \bibfield  {author} {\bibinfo {author} {\bibfnamefont {M.~J.}\ \bibnamefont
  {Hartmann}}, \bibinfo {author} {\bibfnamefont {F.~G. S.~L.}\ \bibnamefont
  {Brandao}}, \ and\ \bibinfo {author} {\bibfnamefont {M.~B.}\ \bibnamefont
  {Plenio}},\ }\bibfield  {title} {\enquote {\bibinfo {title} {Strongly
  interacting polaritons in coupled arrays of cavities},}\ }\href {\doibase
  10.1038/nphys462} {\bibfield  {journal} {\bibinfo  {journal} {Nature
  Physics}\ }\textbf {\bibinfo {volume} {2}},\ \bibinfo {pages} {849--855}
  (\bibinfo {year} {2006})}\BibitemShut {NoStop}%
\bibitem [{\citenamefont {Hartmann}\ \emph {et~al.}(2008)\citenamefont
  {Hartmann}, \citenamefont {Brandao},\ and\ \citenamefont
  {Plenio}}]{hartmann2008quantum}%
  \BibitemOpen
  \bibfield  {author} {\bibinfo {author} {\bibfnamefont {M.~J.}\ \bibnamefont
  {Hartmann}}, \bibinfo {author} {\bibfnamefont {F.~G. S.~L.}\ \bibnamefont
  {Brandao}}, \ and\ \bibinfo {author} {\bibfnamefont {M.~B.}\ \bibnamefont
  {Plenio}},\ }\bibfield  {title} {\enquote {\bibinfo {title} {Quantum
  many-body phenomena in coupled cavity arrays},}\ }\href {\doibase
  10.1002/lpor.200810046} {\bibfield  {journal} {\bibinfo  {journal} {Laser \&
  Photonics Reviews}\ }\textbf {\bibinfo {volume} {2}},\ \bibinfo {pages}
  {527--556} (\bibinfo {year} {2008})}\BibitemShut {NoStop}%
\bibitem [{\citenamefont {Lei}\ and\ \citenamefont
  {Lee}(2008)}]{lei2008quantum}%
  \BibitemOpen
  \bibfield  {author} {\bibinfo {author} {\bibfnamefont {S.~C.}\ \bibnamefont
  {Lei}}\ and\ \bibinfo {author} {\bibfnamefont {R.~K.}\ \bibnamefont {Lee}},\
  }\bibfield  {title} {\enquote {\bibinfo {title} {Quantum phase transitions of
  light in the dicke-bose-hubbard model},}\ }\href {\doibase
  10.1103/PhysRevA.77.033827} {\bibfield  {journal} {\bibinfo  {journal}
  {Physical Review A—Atomic, Molecular, and Optical Physics}\ }\textbf
  {\bibinfo {volume} {77}},\ \bibinfo {pages} {033827} (\bibinfo {year}
  {2008})}\BibitemShut {NoStop}%
\bibitem [{\citenamefont {Lu}\ and\ \citenamefont
  {Wang}(2016{\natexlab{a}})}]{lu2016influencef}%
  \BibitemOpen
  \bibfield  {author} {\bibinfo {author} {\bibfnamefont {Y.}~\bibnamefont
  {Lu}}\ and\ \bibinfo {author} {\bibfnamefont {C.}~\bibnamefont {Wang}},\
  }\bibfield  {title} {\enquote {\bibinfo {title} {Influence of
  counter-rotating interaction on quantum phase transition in {Dicke-Hubbard}
  lattice: an extended coherent-state approach},}\ }\href {\doibase
  10.1007/s11128-016-1392-y} {\bibfield  {journal} {\bibinfo  {journal}
  {Quantum Information Processing}\ }\textbf {\bibinfo {volume} {15}},\
  \bibinfo {pages} {4347--4359} (\bibinfo {year}
  {2016}{\natexlab{a}})}\BibitemShut {NoStop}%
\bibitem [{\citenamefont {Kirton}\ \emph {et~al.}(2019)\citenamefont {Kirton},
  \citenamefont {Roses}, \citenamefont {Keeling},\ and\ \citenamefont
  {Torre}}]{kirton2019introduction}%
  \BibitemOpen
  \bibfield  {author} {\bibinfo {author} {\bibfnamefont {P.}~\bibnamefont
  {Kirton}}, \bibinfo {author} {\bibfnamefont {M.~M.}\ \bibnamefont {Roses}},
  \bibinfo {author} {\bibfnamefont {J.}~\bibnamefont {Keeling}}, \ and\
  \bibinfo {author} {\bibfnamefont {E.~G.~Dalla}\ \bibnamefont {Torre}},\
  }\bibfield  {title} {\enquote {\bibinfo {title} {Introduction to the dicke
  model: From equilibrium to nonequilibrium, and vice versa (adv. quantum
  technol. 1-2/2019)},}\ }\href {\doibase 10.1002/qute.201970013} {\bibfield
  {journal} {\bibinfo  {journal} {Advanced Quantum Technologies}\ }\textbf
  {\bibinfo {volume} {2}},\ \bibinfo {pages} {1970013} (\bibinfo {year}
  {2019})}\BibitemShut {NoStop}%
\bibitem [{\citenamefont {Schir{\'o}}\ \emph {et~al.}(2013)\citenamefont
  {Schir{\'o}}, \citenamefont {Bordyuh}, \citenamefont {{\"O}ztop},\ and\
  \citenamefont {T{\"u}reci}}]{schiro2013quantum}%
  \BibitemOpen
  \bibfield  {author} {\bibinfo {author} {\bibfnamefont {M.}~\bibnamefont
  {Schir{\'o}}}, \bibinfo {author} {\bibfnamefont {M.}~\bibnamefont {Bordyuh}},
  \bibinfo {author} {\bibfnamefont {B.}~\bibnamefont {{\"O}ztop}}, \ and\
  \bibinfo {author} {\bibfnamefont {H.~E.}\ \bibnamefont {T{\"u}reci}},\
  }\bibfield  {title} {\enquote {\bibinfo {title} {Quantum phase transition of
  light in the rabi--hubbard model},}\ }\href {\doibase
  10.1088/0953-4075/46/22/224021} {\bibfield  {journal} {\bibinfo  {journal}
  {Journal of Physics B: Atomic, Molecular and Optical Physics}\ }\textbf
  {\bibinfo {volume} {46}},\ \bibinfo {pages} {224021} (\bibinfo {year}
  {2013})}\BibitemShut {NoStop}%
\bibitem [{\citenamefont {Lu}\ and\ \citenamefont
  {Wang}(2016{\natexlab{b}})}]{lu2016influence}%
  \BibitemOpen
  \bibfield  {author} {\bibinfo {author} {\bibfnamefont {Y.}~\bibnamefont
  {Lu}}\ and\ \bibinfo {author} {\bibfnamefont {C.}~\bibnamefont {Wang}},\
  }\bibfield  {title} {\enquote {\bibinfo {title} {Influence of
  counter-rotating interaction on quantum phase transition in dicke-hubbard
  lattice: an extended coherent-state approach},}\ }\href {\doibase
  10.1007/s11128-016-1373-1} {\bibfield  {journal} {\bibinfo  {journal}
  {Quantum Information Processing}\ }\textbf {\bibinfo {volume} {15}},\
  \bibinfo {pages} {4347--4359} (\bibinfo {year}
  {2016}{\natexlab{b}})}\BibitemShut {NoStop}%
\bibitem [{\citenamefont {Chen}\ \emph {et~al.}(2008)\citenamefont {Chen},
  \citenamefont {Zhang}, \citenamefont {Liu},\ and\ \citenamefont
  {Wang}}]{chen2008numerically}%
  \BibitemOpen
  \bibfield  {author} {\bibinfo {author} {\bibfnamefont {Q.~H.}\ \bibnamefont
  {Chen}}, \bibinfo {author} {\bibfnamefont {Y.~Y.}\ \bibnamefont {Zhang}},
  \bibinfo {author} {\bibfnamefont {T.}~\bibnamefont {Liu}}, \ and\ \bibinfo
  {author} {\bibfnamefont {K.~L.}\ \bibnamefont {Wang}},\ }\bibfield  {title}
  {\enquote {\bibinfo {title} {Numerically exact solution to the finite-size
  dicke model},}\ }\href {\doibase 10.1103/PhysRevA.78.051801} {\bibfield
  {journal} {\bibinfo  {journal} {Physical Review A}\ }\textbf {\bibinfo
  {volume} {78}},\ \bibinfo {pages} {051801} (\bibinfo {year}
  {2008})}\BibitemShut {NoStop}%
\bibitem [{\citenamefont {Chen}\ \emph {et~al.}(2010)\citenamefont {Chen},
  \citenamefont {Liu}, \citenamefont {Zhang},\ and\ \citenamefont
  {Wang}}]{chen2010quantum}%
  \BibitemOpen
  \bibfield  {author} {\bibinfo {author} {\bibfnamefont {Q.~H.}\ \bibnamefont
  {Chen}}, \bibinfo {author} {\bibfnamefont {T.}~\bibnamefont {Liu}}, \bibinfo
  {author} {\bibfnamefont {Y.~Y.}\ \bibnamefont {Zhang}}, \ and\ \bibinfo
  {author} {\bibfnamefont {K.~L.}\ \bibnamefont {Wang}},\ }\bibfield  {title}
  {\enquote {\bibinfo {title} {Quantum phase transitions in coupled two-level
  atoms in a single-mode cavity},}\ }\href {\doibase
  10.1103/PhysRevA.82.053841} {\bibfield  {journal} {\bibinfo  {journal}
  {Physical Review A}\ }\textbf {\bibinfo {volume} {82}},\ \bibinfo {pages}
  {053841} (\bibinfo {year} {2010})}\BibitemShut {NoStop}%
\bibitem [{\citenamefont {Ye}\ \emph {et~al.}(2021)\citenamefont {Ye},
  \citenamefont {Wang},\ and\ \citenamefont {Chen}}]{ye2021quantum}%
  \BibitemOpen
  \bibfield  {author} {\bibinfo {author} {\bibfnamefont {T.}~\bibnamefont
  {Ye}}, \bibinfo {author} {\bibfnamefont {C.}~\bibnamefont {Wang}}, \ and\
  \bibinfo {author} {\bibfnamefont {Q.~H.}\ \bibnamefont {Chen}},\ }\bibfield
  {title} {\enquote {\bibinfo {title} {Quantum phase transition of light in the
  dissipative rabi-hubbard lattice: A dressed-master-equation perspective},}\
  }\href {\doibase 10.1103/PhysRevA.104.053708} {\bibfield  {journal} {\bibinfo
   {journal} {Physical Review A}\ }\textbf {\bibinfo {volume} {104}},\ \bibinfo
  {pages} {053708} (\bibinfo {year} {2021})}\BibitemShut {NoStop}%
\bibitem [{\citenamefont {Boit{\'e}}\ \emph {et~al.}(2016)\citenamefont
  {Boit{\'e}}, \citenamefont {Hwang}, \citenamefont {Nha},\ and\ \citenamefont
  {Plenio}}]{le2016fate}%
  \BibitemOpen
  \bibfield  {author} {\bibinfo {author} {\bibfnamefont {A.~Le}\ \bibnamefont
  {Boit{\'e}}}, \bibinfo {author} {\bibfnamefont {M.~J.}\ \bibnamefont
  {Hwang}}, \bibinfo {author} {\bibfnamefont {H.}~\bibnamefont {Nha}}, \ and\
  \bibinfo {author} {\bibfnamefont {M.~B.}\ \bibnamefont {Plenio}},\ }\bibfield
   {title} {\enquote {\bibinfo {title} {Fate of photon blockade in the deep
  strong-coupling regime},}\ }\href {\doibase 10.1103/PhysRevA.94.033827}
  {\bibfield  {journal} {\bibinfo  {journal} {Physical Review A}\ }\textbf
  {\bibinfo {volume} {94}},\ \bibinfo {pages} {033827} (\bibinfo {year}
  {2016})}\BibitemShut {NoStop}%
\bibitem [{\citenamefont {Settineri}\ \emph {et~al.}(2018)\citenamefont
  {Settineri}, \citenamefont {Macr{\'\i}}, \citenamefont {Ridolfo},
  \citenamefont {Stefano}, \citenamefont {Kockum}, \citenamefont {Nori},\ and\
  \citenamefont {Savasta}}]{settineri2018dissipation}%
  \BibitemOpen
  \bibfield  {author} {\bibinfo {author} {\bibfnamefont {A.}~\bibnamefont
  {Settineri}}, \bibinfo {author} {\bibfnamefont {V.}~\bibnamefont
  {Macr{\'\i}}}, \bibinfo {author} {\bibfnamefont {A.}~\bibnamefont {Ridolfo}},
  \bibinfo {author} {\bibfnamefont {O.~Di}\ \bibnamefont {Stefano}}, \bibinfo
  {author} {\bibfnamefont {A.~F.}\ \bibnamefont {Kockum}}, \bibinfo {author}
  {\bibfnamefont {F.}~\bibnamefont {Nori}}, \ and\ \bibinfo {author}
  {\bibfnamefont {S.}~\bibnamefont {Savasta}},\ }\bibfield  {title} {\enquote
  {\bibinfo {title} {Dissipation and thermal noise in hybrid quantum systems in
  the ultrastrong-coupling regime},}\ }\href {\doibase
  10.1103/PhysRevA.98.053834} {\bibfield  {journal} {\bibinfo  {journal}
  {Physical Review A}\ }\textbf {\bibinfo {volume} {98}},\ \bibinfo {pages}
  {053834} (\bibinfo {year} {2018})}\BibitemShut {NoStop}%
\bibitem [{\citenamefont {Beaudoin}\ \emph {et~al.}(2011)\citenamefont
  {Beaudoin}, \citenamefont {Gambetta},\ and\ \citenamefont
  {Blais}}]{beaudoin2011dissipation}%
  \BibitemOpen
  \bibfield  {author} {\bibinfo {author} {\bibfnamefont {F.}~\bibnamefont
  {Beaudoin}}, \bibinfo {author} {\bibfnamefont {J.~M.}\ \bibnamefont
  {Gambetta}}, \ and\ \bibinfo {author} {\bibfnamefont {A.}~\bibnamefont
  {Blais}},\ }\bibfield  {title} {\enquote {\bibinfo {title} {Dissipation and
  ultrastrong coupling in circuit qed},}\ }\href {\doibase
  10.1103/PhysRevA.84.043832} {\bibfield  {journal} {\bibinfo  {journal}
  {Physical Review A}\ }\textbf {\bibinfo {volume} {84}},\ \bibinfo {pages}
  {043832} (\bibinfo {year} {2011})}\BibitemShut {NoStop}%
\bibitem [{\citenamefont {Gardiner}\ and\ \citenamefont
  {Collett}(1985)}]{gardiner1985input}%
  \BibitemOpen
  \bibfield  {author} {\bibinfo {author} {\bibfnamefont {C.~W.}\ \bibnamefont
  {Gardiner}}\ and\ \bibinfo {author} {\bibfnamefont {M.~J.}\ \bibnamefont
  {Collett}},\ }\bibfield  {title} {\enquote {\bibinfo {title} {Input and
  output in damped quantum systems: Quantum stochastic differential equations
  and the master equation},}\ }\href {\doibase 10.1103/PhysRevA.31.3761}
  {\bibfield  {journal} {\bibinfo  {journal} {Physical Review A}\ }\textbf
  {\bibinfo {volume} {31}},\ \bibinfo {pages} {3761} (\bibinfo {year}
  {1985})}\BibitemShut {NoStop}%
\bibitem [{\citenamefont {Quan}\ \emph
  {et~al.}(2007{\natexlab{b}})\citenamefont {Quan}, \citenamefont {Liu},
  \citenamefont {Sun},\ and\ \citenamefont {Nori}}]{quan2007quantum}%
  \BibitemOpen
  \bibfield  {author} {\bibinfo {author} {\bibfnamefont {H.~T.}\ \bibnamefont
  {Quan}}, \bibinfo {author} {\bibfnamefont {Y.~X.}\ \bibnamefont {Liu}},
  \bibinfo {author} {\bibfnamefont {C.~P.}\ \bibnamefont {Sun}}, \ and\
  \bibinfo {author} {\bibfnamefont {F.}~\bibnamefont {Nori}},\ }\bibfield
  {title} {\enquote {\bibinfo {title} {Quantum thermodynamic cycles and quantum
  heat engines},}\ }\href {\doibase 10.1103/PhysRevE.76.031105} {\bibfield
  {journal} {\bibinfo  {journal} {Physical Review E}\ }\textbf {\bibinfo
  {volume} {76}},\ \bibinfo {pages} {031105} (\bibinfo {year}
  {2007}{\natexlab{b}})}\BibitemShut {NoStop}%
\bibitem [{\citenamefont {Quan}(2009)}]{quan2009quantum}%
  \BibitemOpen
  \bibfield  {author} {\bibinfo {author} {\bibfnamefont {H.~T.}\ \bibnamefont
  {Quan}},\ }\bibfield  {title} {\enquote {\bibinfo {title} {Quantum
  thermodynamic cycles and quantum heat engines. ii.}}\ }\href {\doibase
  10.1103/PhysRevE.79.041129} {\bibfield  {journal} {\bibinfo  {journal}
  {Physical Review E}\ }\textbf {\bibinfo {volume} {79}},\ \bibinfo {pages}
  {041129} (\bibinfo {year} {2009})}\BibitemShut {NoStop}%
\bibitem [{\citenamefont {Kieu}(2004)}]{kieu2004second}%
  \BibitemOpen
  \bibfield  {author} {\bibinfo {author} {\bibfnamefont {T.~D.}\ \bibnamefont
  {Kieu}},\ }\bibfield  {title} {\enquote {\bibinfo {title} {The second law,
  maxwell's demon, and work derivable from quantum heat engines},}\ }\href
  {\doibase 10.1103/PhysRevLett.93.140403} {\bibfield  {journal} {\bibinfo
  {journal} {Physical review letters}\ }\textbf {\bibinfo {volume} {93}},\
  \bibinfo {pages} {140403} (\bibinfo {year} {2004})}\BibitemShut {NoStop}%
\bibitem [{\citenamefont {Solfanelli}\ \emph {et~al.}(2020)\citenamefont
  {Solfanelli}, \citenamefont {Falsetti},\ and\ \citenamefont
  {Campisi}}]{solfanelli2020nonadiabatic}%
  \BibitemOpen
  \bibfield  {author} {\bibinfo {author} {\bibfnamefont {A.}~\bibnamefont
  {Solfanelli}}, \bibinfo {author} {\bibfnamefont {M.}~\bibnamefont
  {Falsetti}}, \ and\ \bibinfo {author} {\bibfnamefont {M.}~\bibnamefont
  {Campisi}},\ }\bibfield  {title} {\enquote {\bibinfo {title} {Nonadiabatic
  single-qubit quantum otto engine},}\ }\href {\doibase
  10.1103/PhysRevB.101.054513} {\bibfield  {journal} {\bibinfo  {journal}
  {Physical Review B}\ }\textbf {\bibinfo {volume} {101}},\ \bibinfo {pages}
  {054513} (\bibinfo {year} {2020})}\BibitemShut {NoStop}%
\end{thebibliography}%

\end{document}